\PassOptionsToPackage{table,xcdraw}{xcolor}

\documentclass[sigconf,nonacm]{acmart}
\AtBeginDocument{%
  \providecommand\BibTeX{{%
    \normalfont B\kern-0.5em{\scshape i\kern-0.25em b}\kern-0.8em\TeX}}}

\usepackage{graphicx}
\usepackage{caption}
\usepackage{multirow}
\usepackage{bigstrut}
\usepackage{cellspace}
\usepackage{subfigure}
\usepackage{float}
\usepackage{placeins}

\begin{document}

\title[Design and Usage of an AI-Powered Video-Based American Sign Language Dictionary
]{Towards an AI-Driven Video-Based American Sign Language Dictionary: Exploring Design and Usage Experience with Learners}


\author{Saad Hassan}
\affiliation{
  \institution{Tulane University}
  \city{New Orleans}
  \state{LA}
  \country{USA}}
\email{saadhassan@tulane.edu}

\author{Matyas Bohacek}
\affiliation{
  \institution{Stanford University}
  \city{Stanford}
  \state{CA}
  \country{USA}}
\email{maty@stanford.edu}

\author{Chaelin Kim}
\affiliation{
  \institution{Tulane University}
  \city{New Orleans}
  \state{LA}
  \country{USA}}
\email{ckim13@tulane.edu}

\author{Denise Crochet}
\affiliation{
  \institution{Tulane University}
  \city{New Orleans}
  \state{LA}
  \country{USA}}
\email{crochet@tulane.edu}

\renewcommand{\shortauthors}{Hassan, et al.}

\begin{abstract}

Searching for unfamiliar American Sign Language (ASL) signs is challenging for learners because, unlike spoken languages, they cannot type a text-based query to look up an unfamiliar sign. Advances in isolated sign recognition have enabled the creation of video-based dictionaries, allowing users to submit a video and receive a list of the closest matching signs. Previous HCI research using Wizard-of-Oz prototypes has explored interface designs for ASL dictionaries. Building on these studies, we incorporate their design recommendations and leverage state-of-the-art sign-recognition technology to develop an automated video-based dictionary. We also present findings from an observational study with twelve novice ASL learners who used this dictionary during video-comprehension and question-answering tasks. Our results address human-AI interaction challenges not covered in previous WoZ research, including recording and resubmitting signs, unpredictable outputs, system latency, and privacy concerns. These insights offer guidance for designing and deploying video-based ASL dictionary systems.


\end{abstract}





\maketitle

\section{Introduction}\label{introduction}


Learning ASL benefits both Deaf and hard of hearing (DHH) individuals, as well as hearing people, by fostering interaction and promoting societal inclusion \cite{quinto2011teaching}. Various groups, including parents, relatives, and caregivers of DHH people, as well as researchers working with DHH people, are motivated to learn sign language \cite{hall2017language,weaver2011we, schnepp2020human}. There is also a significant demand for basic ASL communication skills in roles that require interactions with DHH people (e.g., healthcare providers, receptionists, and other employment contexts involving public interaction)~\cite{mckee2015emergency,rotoli2022emergency,terry2024scoping}. This overall growing interest in learning ASL is evidenced by an increasing number of students currently enrolled in ASL courses at educational institutions across the U.S. \cite{goldberg2015enrollments,NCED}.



Dictionaries are essential tools for ASL learning and play a crucial role in enhancing the employability of individuals who use ASL in professional contexts that require regular interactions with DHH people. Traditional online ASL dictionaries use feature-based approaches, where users search and filter signs by linguistic properties like handshape, movement, and location \cite{lapiak}. This approach allows users to input only the features they recall, offers minimal latency, and avoids video data storage, reducing privacy concerns. However, it can be difficult for novice learners who may struggle to recall and input linguistic features \cite{hoffmeister2000piece, athitsos2010large, components, bragg2015user}. Identifying sign boundaries and decomposing signs from memory is challenging, leading many learners to abandon searches quickly \cite{10.1145/3491102.3501986, bragg2015user}.

Recent advancements in sign recognition technology have significantly improved the accuracy of isolated sign recognition models \cite{camgoz2020sign, camgoz2018neural, pu2018dilated, RASTGOO2021113794}. These models can potentially support the design of more functional video-based ASL dictionaries, capable of taking a video input of a sign performance and producing a list of matching signs. Sign language recognition approaches have also become more privacy-preserving, as there is no need to store raw videos recorded from users to provide matching signs~\cite{rust2024towards,hameed2022privacy}. Additionally, significant progress has been made in other AI-powered systems that can provide users with feedback on the quality of their video submissions, except for the quality of the actual sign performance~\cite{huenerfauth2017evaluation,paudyal2019learn2sign}. These advancements could enable the design of an accurate, privacy-preserving, low-latency ASL dictionary that can be effectively used by ASL learners in an educational context.

As isolated sign recognition has improved, HCI researchers have started examining ASL dictionary design and ASL learners’ interaction with Wizard-of-Oz (WoZ) prototypes. Given the imperfections in sign recognition, ASL learners still have to sift through extensive results, making efficient navigation essential. Therefore, prior WoZ studies have focused on search result composition and presentation \cite{alonzo2019effect, hassan2020effect, TACCESS, 10.1145/3517428.3544883}.  Despite these challenges, one of these studies showed that ASL learners prefer video-based dictionaries over feature-based ones like the Handspeak Reverse Dictionary \cite{10.1145/3517428.3544883}.

While WoZ studies provide insights into design choices and querying approaches, they leave key questions unanswered for real-world deployment. These studies typically use isolated video stimuli, asking users to record signs immediately \cite{10.1145/3491102.3501986} or segment clips from provided videos \cite{10.1145/3517428.3544883}. They do not account for users reattempting signs from memory, making recording errors, or refining inputs to improve search results. They also overlook post-query steps, such as acceptable latency after submission. The WoZ prototypes used in prior research were agnostic to input, merely simulating the recognition model’s performance with pre-curated results \cite{10.1145/3491102.3501986, alonzo2019effect}. As a result, users’ navigation of imperfect outputs, including cases where the desired result is absent, could not be observed.

A recent study introduced a video-based ASL dictionary using a recent sign recognition model and reported preliminary findings from interviews with four ASL learners \cite{bohacek2023sign}. Building on this, we developed an improved ASL dictionary prototype with enhanced feedback on video submissions, latency and confidence displays, and improved privacy features (summarized in Section \ref{design}).

The \textbf{contributions}~\cite{10.1145/2907069} of our work are twofold: 

\begin{enumerate}
\item We present a fully automated ASL dictionary prototype using state-of-the-art sign recognition, privacy-preserving video handling, and video quality feedback. It integrates prior HCI research on system status, confidence levels, and result navigation. The prototype is open-sourced\footnote{\url{https://github.com/matyasbohacek/sign-language-dictionary}}.

\item We present findings from an analysis of observational data and interviews conducted with 12 ASL learners. Our analysis highlights the benefits and usage patterns of our tool during a video comprehension task and uncovers challenges specific to a fully functional dictionary. Additionally, the findings offer feedback on our prototype, focusing on novel features.


\end{enumerate}

\section{Background and Related Work}\label{related-work}


\subsection{Sign Language Dictionaries}

An open-ended survey with ASL learners regarding their use of ASL dictionaries highlighted the need for a user-friendly and reliable web-based tool for ASL sign lookup, revealing that existing resources for finding the English translation of ASL words were rarely used \cite{bragg2015user}. Current ASL dictionaries employ various input modalities for sign language lookup, which can be broadly categorized into two categories: \emph{search-by-feature} (feature-based) and \emph{search-by-video} (video-based).

Several \textit{search-by-feature} ASL dictionaries are commercially available \cite{lapiak, slinto} or have been proposed by researchers \cite{sign_bsl,fuertes2006bilingual,10.1145/1229390.1229401}. As mentioned in the introduction, these dictionaries let users filter signs based on one or more linguistic features. Unlike video-based dictionaries, they do not require video submissions, eliminating privacy concerns and latency issues. However, these dictionaries demand that novice learners decompose the sign they encountered into linguistic features of ASL, often using standard formats like Stokoe notation \cite{mccarty2004notation}. This process can be difficult for novice learners, as they may struggle to recall and break down a sign into its components. Unlike written languages, which allow for spelling approximations, ASL's visual nature makes this approach ineffective, requiring learners to input the correct features to look up signs. Research shows that ASL learners rarely use feature-based dictionaries for looking up unfamiliar signs \cite{bragg2015user}. 

Recent advances in sign recognition technology \cite{camgoz2020sign, elliott2011search, yanovich2016detection, 10.5555/1880751.1880787, camgoz2018neural, pu2018dilated, RASTGOO2021113794} and the availability of large-scale publicly accessible datasets, such as ASL Citizen \cite{desai2024asl}\footnote{\url{https://www.microsoft.com/en-us/research/project/asl-citizen/datasheet/}}, ASLLVD \cite{Neidle_2012_Challenges}\footnote{\url{http://dai.cs.rutgers.edu/dai/s/signbank}}, and PopSign \cite{NEURIPS2023_00dada60}\footnote{\url{https://signdata.cc.gatech.edu/view/datasets/popsign_v1_0/index.html}}, have facilitated the development of \textit{search-by-video} ASL dictionaries, allowing users to submit video samples without selecting specific linguistic features. These dictionaries are particularly useful when a user has a video of the unknown sign or can easily recall its performance from memory. However, challenges remain, as recognizing 3D signs from 2D videos is still difficult, and factors like poor video quality or background noise further hinder recognition accuracy. One study found that the correct sign appeared in the top 20 search results up to 67\% of the time for a limited set of signs \cite{athitsos2010large}, indicating that users may need to browse through results or re-record their videos to find the correct sign. While the quality of these models has improved \cite{bohavcek2022sign}, they are not yet perfected, and challenges related to HAI—such as recording a high-quality video—persist. In our work, we present design solutions to address these challenges using AI techniques, and we seek user feedback on these features through our observational study.

\begin{figure*}[t]
    \centering
    \subfigure[Movement]{
        \includegraphics[width=0.175\linewidth]{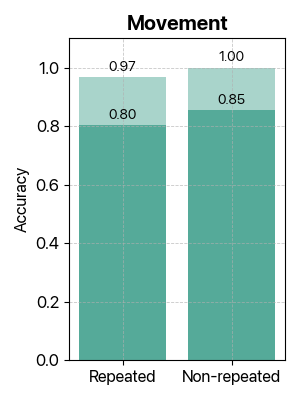}
        \label{fig:movement}
    }
    \hfill
    \subfigure[Number of Hands]{
        \includegraphics[width=0.175\linewidth]{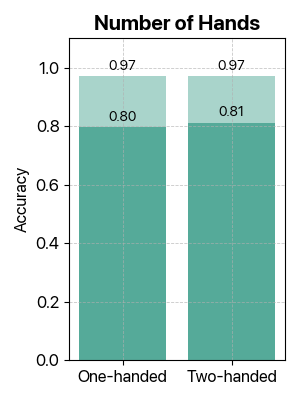}
        \label{fig:number-of-hands}
    }
    \hfill
    \subfigure[Location]{
        \includegraphics[width=0.345\linewidth]{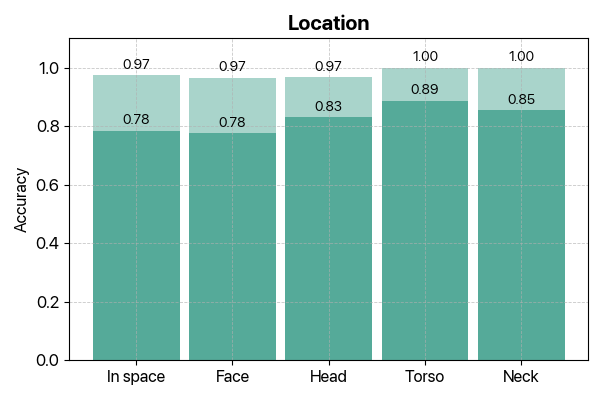}
        \label{fig:location}
    }
    \hfill
    \subfigure{
        \includegraphics[width=0.120\linewidth]{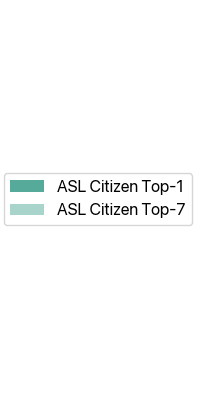}
    }
    
    \caption{Performance of our sign language recognition model for signs grouped by movement (a), number of hands (b), and location (c). For each feature, top-1 and top-7 accuracies on ASL Citizen are reported. \emph{ASL Citizen} top-1 accuracy is shown in vivid green \textcolor[HTML]{5aa99f}{$\blacksquare$}, while top-7 appears in vivid green with half opacity \textcolor[HTML]{add4cf}{$\blacksquare$}.}
    \Description{The figure shows three bar plots, laid out into two rows, each corresponding to a specific feature of signs and their testing accuracy. Each plot contains labeled bars where darker sections represent Top-1 accuracy and lighter extensions represent Top-7 accuracy.}
    \label{fig:feature-level-accuracy}
\end{figure*}

\subsection{HCI Research on ASL Dictionaries}\label{hci-research}

HCI researchers have explored various aspects of ASL dictionaries, such as the design of the search results page, navigation through results, and the ease of narrowing down searches \cite{majetic2017proposing, 10.1145/3491102.3501986}. Notable findings include providing significant linguistic information in results, using video results, offering post-query refinement options, and conveying recognition performance confidence \cite{10.1145/3491102.3501986, bohacek2023sign}. A recent prior work on video-based dictionaries compared different querying approaches for ASL dictionaries \cite{10.1145/3491102.3501986} and found that methods combining search-by-video with search-by-feature as a post-query refinement step yield the highest user satisfaction. We incorporate these design recommendations into our video-based dictionary prototype described in more detail in Section~\ref{design}. 

Researchers have also explored how the presentation and accuracy of search results impact user satisfaction \cite{alonzo2019effect, hassan2020effect, TACCESS}, and have proposed metrics to evaluate sign recognition performance. These metrics extend traditional search metrics, like Discounted Cumulative Gain (DCG) from information retrieval literature, adapting them to the search-by-video context in ASL dictionaries. Notably, one metric combines the placement of the desired result with the precision of other results to provide a more comprehensive evaluation of model performance. In describing our prototype, we assess its performance using one of the metrics proposed in prior research \cite{TACCESS}. We also report additional performance metrics relevant to a functional video-based dictionary, such as performance across various linguistic types of signs (Section \ref{feature-performance}) and the impact of varying video submission quality (Section \ref{resolution-performance}).

Prior HCI research has revealed the benefits of video-based dictionaries \cite{10.1145/3517428.3544883}, but their experimental conditions lacked full ecological validity. Users typically performed isolated signs in front of a webcam immediately after viewing them. In real-life scenarios, an ASL learner might use such a dictionary to look up an unfamiliar sign during a course assignment, after watching a video, or following a signed conversation. These situations make it significantly more challenging for novice learners, as they must isolate a sign from its context (either from memory or a video) and attempt to replicate it, which can result in more unpredictable model performance. Our observational study used a fully functional ASL dictionary prototype to explore these challenges in a class exercise usage context.

\subsection{Human-AI Interaction with Video Systems}\label{HAI}


Matching a user's video to a database is imperfect, often requiring users to browse multiple results. Effective presentation is crucial, as users may struggle to find the correct match. While ``search-by-performance'' removes the need for text queries or sample videos, it still relies on accurate recall from memory. These challenges highlight the need for \emph{providing users feedback on their video queries} and \emph{conveying confidence in the results}. Though underexplored in search-by-video dictionaries, insights from HAI and information retrieval, such as ``search-by-image'' or text-based search, may help.

Researchers have explored methods for providing users with feedback on query quality. Systems automatically assess factors like lighting and framing, prompting users to resubmit if thresholds are not met \cite{dhole2024queryexplorer, 10.1145/3391613}. AI-based systems, such as ID verification and facial recognition, offer near-real-time feedback to help users refine inputs \cite{wong2010interactive}. Studies also examine user behavior in improving AI performance \cite{daly2021user}, willingness to invest effort when systems respond \cite{dennis2023ai}, and how users assume responsibility for errors \cite{10.1145/3491102.3517565, schoenherr2024ai}. These insights guided our feedback modules (Section~\ref{feedback}) and our study on user behavior in refining video submissions (Section~\ref{reattempting}).

Providing confidence levels for AI predictions improves trust and decision-making \cite{10.1145/3290605.3300233}, helping users decide whether to revise a query or sift through results. Studies show users engage more with systems that offer transparent feedback and respond to adjustments. HAI guidelines recommend graded confidence indicators rather than numerical (e.g., 86\%) or binary (e.g., ``yes/no'') formats and suggest contextualizing confidence with explanations \cite{10.1145/3544548.3581278}. Google's People + AI Research guidelines advise presenting confidence in an interpretable way and offering recourse when confidence is low \cite{google2024pair}. Our prototype ranks results by relevance, includes confidence labels and explanations (Section~\ref{design}), and examines participant interactions with these indicators (Section~\ref{unpredictability}).

\subsection{Research Questions}

Our observational study aims to understand how novice learners use a fully functional video-based ASL dictionary prototype during a video comprehension and question-answering task. We also seek feedback for further improvements.
We have two research questions:
\begin{enumerate}
    \item[RQ1.] How did ASL learners use a  functional video-based ASL dictionary prototype to perform sign searches while answering comprehension questions about ASL videos?
    \item[RQ2.] How do ASL learners believe their experience with the ASL dictionary  can be improved?
\end{enumerate}

\section{Sign Language Dictionary Prototype}\label{design}

We adapted the initial design of our ASL dictionary prototype from recent prior work~\cite{bohacek2023sign}, including the underlying ASL recognition model~\cite{bohacek2022sign}. We introduced several improvements based on the findings outlined in their paper. To support reproducibility and further research, we have open-sourced the prototype\footnote{\url{https://github.com/matyasbohacek/sign-language-dictionary/}}.

\subsection{Underlying ASL Recognition and Feedback Methods}\label{ai-pipeline}

Our ASL dictionary prototype is powered by a custom AI pipeline, providing feedback on the input videos submitted by users and recognizing signs from a vocabulary set. 


\subsubsection{Implementation and Training Details}

We used the Transformer-based SPOTER~\cite{bohacek2022sign} architecture, as suggested by the authors of prior ASL dictionary work~\cite{bohacek2023sign}. The model was implemented in PyTorch~\cite{Paszke2019PyTorchAI} and trained for $100$ epochs using the default hyperparameters\footnote{\textbf{Training hyperparameters:} epochs: $100$, scheduler factor: $0.1$, scheduler patience $5$, learning rate: $E-3$; \textbf{Augmentation hyperparameters:} augmentations probability: $0.5$, max arm joint rotate angle: $4$, arm joint rotate probability: $0.4$, max rotate angle $17$, max squeeze ratio $0.4$, max perspective transform ratio: $0.2$.} reported in the original work~\cite{bohacek2022sign}. Our implementation slightly differs from \cite{bohacek2023sign} in that individual sign renditions were treated as separate classes (as opposed to combining distinct signs with equivalent meanings under one class). As such, this is not an architectural change and is unlikely to impact system performance, except for improving the accuracy of signs with distinct variations.


\begin{figure}[t]
    \centering
    \includegraphics[width=0.9\linewidth]{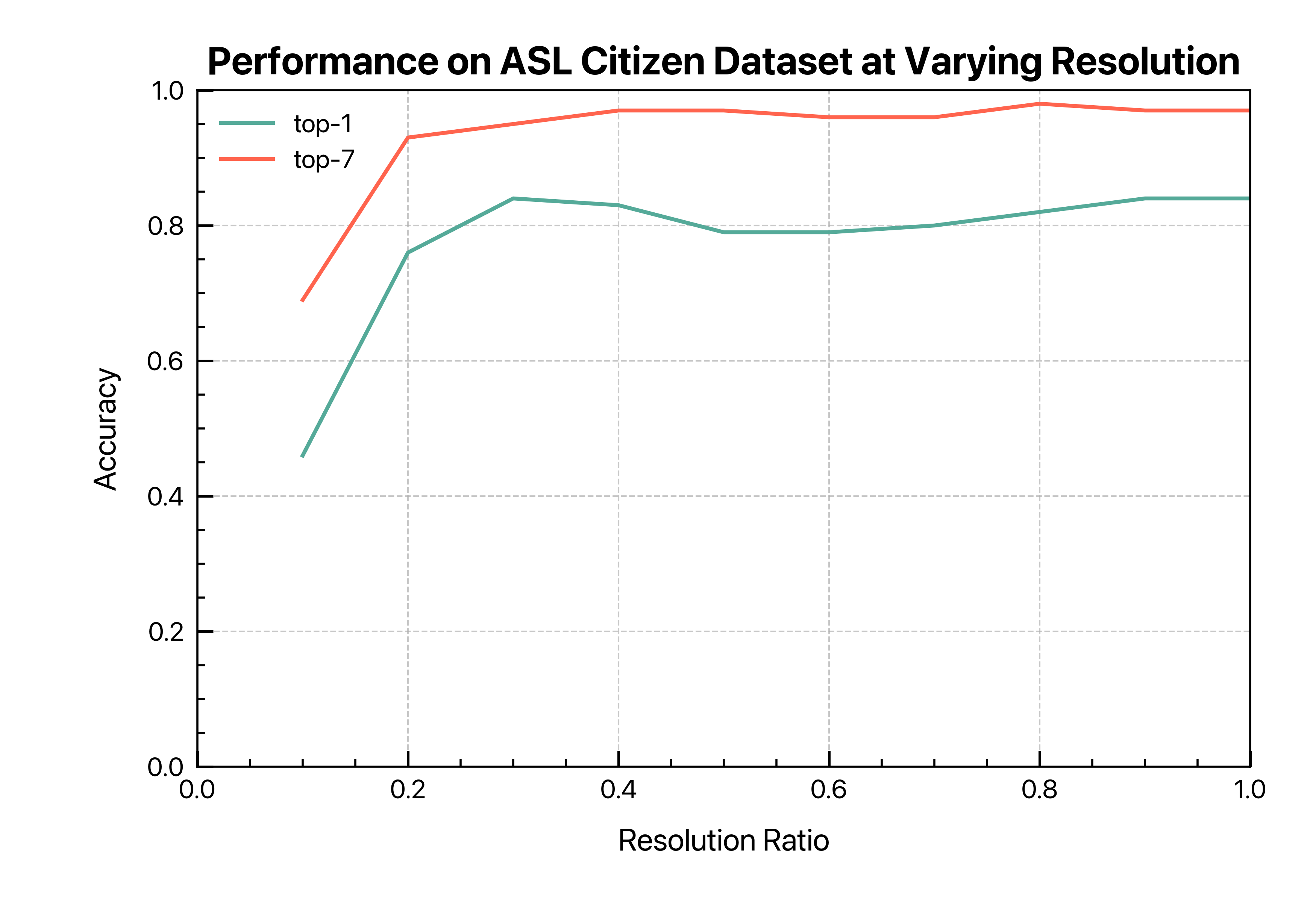}
    \Description{The figure shows two line plots comparing the top-1 and top-7 testing accuracy of our sign language recognition model for ASL Citizen under decreased input video resolution. The x-axis represents the resolution ratio (on the range from 0.0 to 1.0), and the y-axis represents accuracy (on the range from 0.0 to 1.0). The green line represents top-1 accuracy, while the red line represents top-7 accuracy. The top-1 accuracy on ASL Citizen line data points are located at the following coordinates: (0.1, 0.46); (0.2, 0.76); (0.3, 0.84); (0.4, 0.83); (0.5, 0.79); (0.6, 0.79); (0.7, 0.80); (0.8, 0.82); (0.9, 0.84); (1.0, 0.84).}
    \caption{The top-1 and top-7 testing accuracy of our sign language recognition model (y-axis) for input videos of increasing resolution ratio (x-axis) on ASL Citizen. Vivid green \textcolor[HTML]{5aa99f}{$\blacksquare$} represents top-1 accuracy; vivid red \textcolor[HTML]{cc5555}{$\blacksquare$} represents top-7. The resolution ratio indicates video resolution relative to the standard $640\times480$ pixels.}
    \label{fig:resolution-accuracy}
\end{figure}

\subsubsection{Training Data}\label{subsubsec:training-data}

The vocabulary set used for training (described further in Section~\ref{materials}) included signs related to countries, animals, and food. The selection of these signs was done in consultation with the ASL educator on our team and aligned with the ASL II course structure at the university where our observational study was conducted. As noted earlier, we trained the model specifically on one course module's vocabulary to minimize the risk of unexpected performance inconsistencies during the  study.

The ASL Citizen dataset~\cite{desai2024asl}, manually mapped onto our vocabulary set, was used to compile the training set. Additional publicly available videos were used to fill in gaps when the ASL Citizen dataset~\cite{desai2024asl} did not contain a desired sign. This dataset comprises $177$ entries with $146$ unique signs. There are $18$ signs that have multiple versions, meaning there are two or more different performances of the signs mapping to the same word in written language. We followed the train-test split of the dataset. 



\begin{figure}[t]
    \centering
    \includegraphics[width=0.9\linewidth]{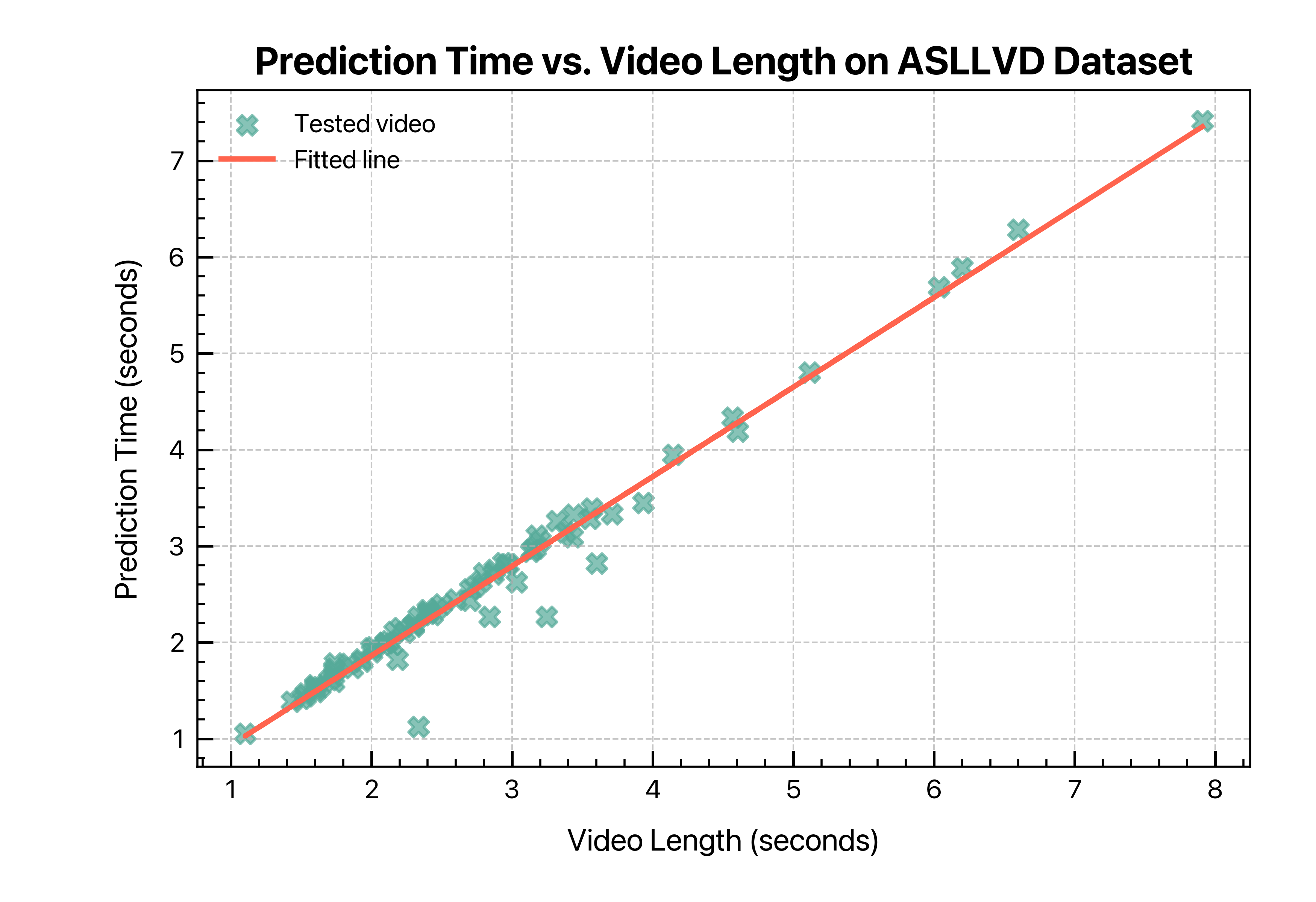}
    \caption{Latency analysis of the custom sign recognition AI model, showing the prediction time as a function of input video length. A linear line is fitted on the data.}
    \Description{The figure shows a scatter plot with markers and a fitted line showing a linear relationship. The x-axis represents the video length in seconds and ranges from 1 to 8. The y-axis represents the prediction time in seconds and ranges from 1 to 7. The figure shows a legend in the top left corner, indicating that data points are scattered as green markers and that the fitted line is orange. The data points are located at the following coordinates: (1.60,1.48); (1.73,1.59); (1.50,1.42); (1.93,1.80); (2.13,2.01); (1.93,1.83); (1.63,1.54); (1.43,1.39); (1.74,1.78); (1.87,1.74); (1.74,1.73); (1.53,1.47); (2.10,1.99); (1.84,1.78); (1.60,1.51); (1.60,1.54); (1.73,1.65); (2.00,1.90); (2.02,1.95); (2.00,1.94); (1.10,1.05); (2.10,2.00); (1.60,1.56); (1.53,1.44); (1.70,1.61); (1.73,1.70); (2.30,2.18); (2.60,2.45); (2.30,2.17); (2.24,2.10); (2.21,2.15); (2.47,2.35); (2.43,2.29); (2.84,2.27); (2.70,2.43); (2.34,2.27); (2.87,2.75); (2.70,2.55); (2.94,2.82); (2.80,2.72); (2.30,2.18); (2.34,1.12); (2.50,2.39); (2.40,2.30); (2.16,2.11); (2.74,2.58); (2.78,2.64); (2.40,2.31); (2.87,2.71); (2.40,2.34); (2.15,1.98); (2.19,1.82); (4.14,3.95); (3.93,3.45); (4.57,4.33); (6.60,6.28); (2.97,2.81); (3.57,3.39); (3.17,3.11); (3.25,2.27); (3.60,2.82); (3.14,2.96); (2.96,2.81); (3.55,3.29); (6.03,5.69); (6.20,5.89); (3.32,3.26); (3.17,2.98); (3.43,3.09); (3.38,3.14); (3.20,3.05); (3.71,3.33); (3.03,2.62); (4.60,4.19); (3.44,3.33); (5.11,4.80); (3.13,2.94); (7.91,7.41)}
    \label{fig:latency-analysis}
\end{figure}

\subsubsection{Performance Evaluation}

\paragraph{Overall Accuracy} The sign recognition model achieved a top-$1$ accuracy of $81 \%$ and a top-$7$ accuracy of $97 \%$ on the testing set of ASL Citizen. The mean per-class top-1 accuracy was $81 \%$ ($\sigma$=$18 \%$). We evaluated the top-$7$ accuracy because our updated interface design presents the user with the top-7 predictions on the `Main dictionary interface' page. This metric thus describes the model's accuracy in presenting the correct prediction somewhere on this first result screen (described in detail in Section~\ref{design}). 

\paragraph{Discounted Cumulative Gain (DCG)} To better contextualize our usage patterns uncovered in the observational study, we also evaluated the DCG of the model as suggested by prior HCI research \cite{alonzo2019effect, TACCESS}. Following the adaptation of the normalized DCG (nDCG) and the Ideal DCG (IDCG) metrics for sign language dictionaries~\cite{10.1145/3470650}, we calculated these metrics for a single prediction following Equations~\ref{eq:idcg} and \ref{eq:ndcg}:

\begin{equation}
IDCG_p = \sum_{i=1}^{p} \frac{2^{rel_i} - 1}{\log_2(\pi(i) + 1)}
\label{eq:idcg}
\end{equation}

\begin{equation}
nDCG_p = \frac{DCG_p}{IDCG_p}
\label{eq:ndcg}
\end{equation}

where $rel_i \in {0, 0.5, 1}$. If the $i$-th predicted sign matched the ground truth, $rel_i=1$; if the $i$-th predicted sign did not match the ground truth but shared at least one attribute with the ground truth (same number of hands, handshape, or movement), $rel_i=0.5$; else, $rel_i=0$. On the testing set of ASL Citizen, our model achieved a mean nDSG of $0.92$. 


\paragraph{Feature-level Accuracy}\label{feature-performance} To better assess the differences in our dictionary's performance across various types of signs, we also conducted an analysis of signs sharing similar linguistic properties. Figure~\ref{fig:feature-level-accuracy} presents the top-1 and top-7 accuracies of our model on the ASL Citizen testing set, grouped by features in the following categories: movement, number of hands, and location. The grouping was done by the expert ASL educator on our team based on criteria specified in prior work \cite{10.1145/3491102.3501986}.

The model’s performance on ASL Citizen appears balanced for movement and number of hands features. In terms of sign location, the model performs best at recognizing signs localized on the torso and on neck with a top-1 accuracy of $89 \%$ and $85 \%$, respectively, and a top-7 accuracy of $100 \%$ for both cases. The performance on signs predominantly ``in space'' and localized on the face pose the biggest challenge to our model with a top-1 accuracy of $78 \%$ and a top-7 accuracy of $97 \%$.

\paragraph{Accuracy on Different Resolutions}\label{resolution-performance} Since our prototype is expected to be operated with a variety of cameras, it was useful to evaluate performance on inputs with varying resolutions. Figure~\ref{fig:resolution-accuracy} shows the model's accuracy on the ASL Citizen testing set, assessing its robustness to lower-quality input videos. 

At $64 \times 48$ pixels (10\% of standard resolution), the model achieves a top-1 accuracy of 46\% on ASL Citizen. The accuracy improves to levels comparable to full-resolution performance at $192 \times 144$ pixels (30\% of standard resolution), where it reaches 84\%. For top-7 accuracy, the model recovers even earlier, achieving 93\% at $128 \times 96$ pixels (20\% of standard resolution).


\begin{figure}[t]
    \centering
    \includegraphics[width=0.9\linewidth]{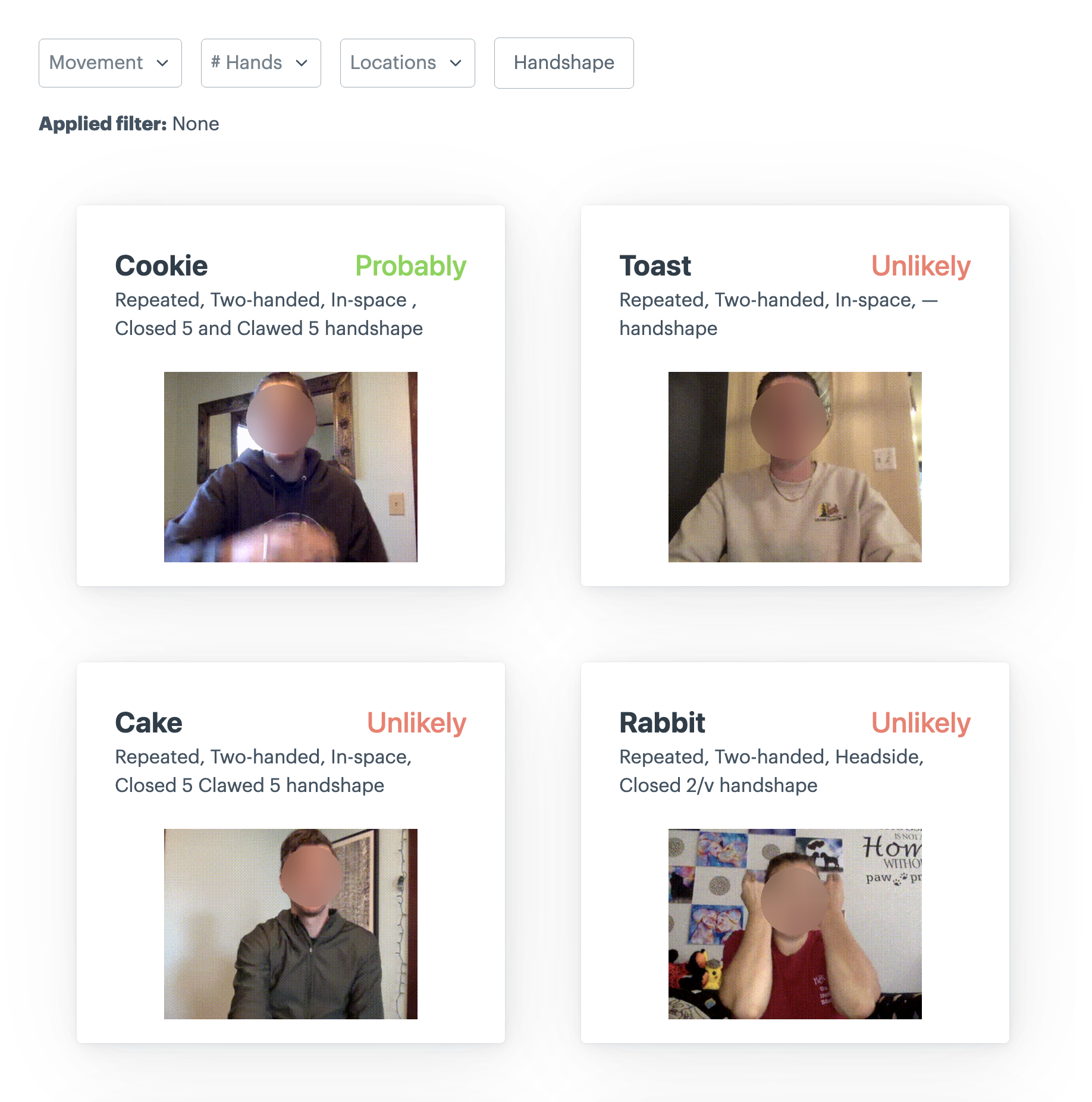}
    \Description['Detailed analysis' results view]{The full-screen view with a white background shows the complete prediction results. Filter buttons with the labels ‘Movement,’ ‘# Hands,’ ‘Locations,’ and ‘Handshape’ are shown at the top left of the view. Below these buttons is a label saying ‘Applied filter: None’. The rest of the view shows a 2x2 grid with prediction items. Each item presents the word translation of the sign on the left and its textual probability on the right. Below these labels is a set of metadata associated with the sign. A blurred-out frame of a video with a person signing the sign is below all these labels. The word translations in the associated items (top-down left-to-right) are ‘Cookie,’ ‘Toast,’ ‘Cake,’ and ‘Rabbit.’}
    \caption{The `Detailed analysis' page presents the model's top $20$ predictions in a scrollable grid of sign entries, ordered by likelihood. Each entry includes a representative recording, word translation, probability score, and metadata.}
    \label{fig:resultsview-detailed}
\end{figure}

These results suggest that the model is robust to video resolution smaller than those seen during training, maintaining stable performance for videos with resolutions larger than $384 \times 288$ pixels---a standard supported by most contemporary webcams. This robustness is likely due to the model's intermediate step of performing pose estimation and processing pose sequences.

\paragraph{Latency} Figure~\ref{fig:latency-analysis} shows an analysis of the sign recognition AI model latency. The analysis was performed on a consumer laptop (2022 MacBook Pro), with the model being inferred on the CPU. It reveals that the time it takes for the model to make predictions increases linearly with the length of the input video. In most cases, the prediction time was shorter than the length of the input video.

\begin{figure}[t]
    \centering
    \includegraphics[width=0.9\linewidth]{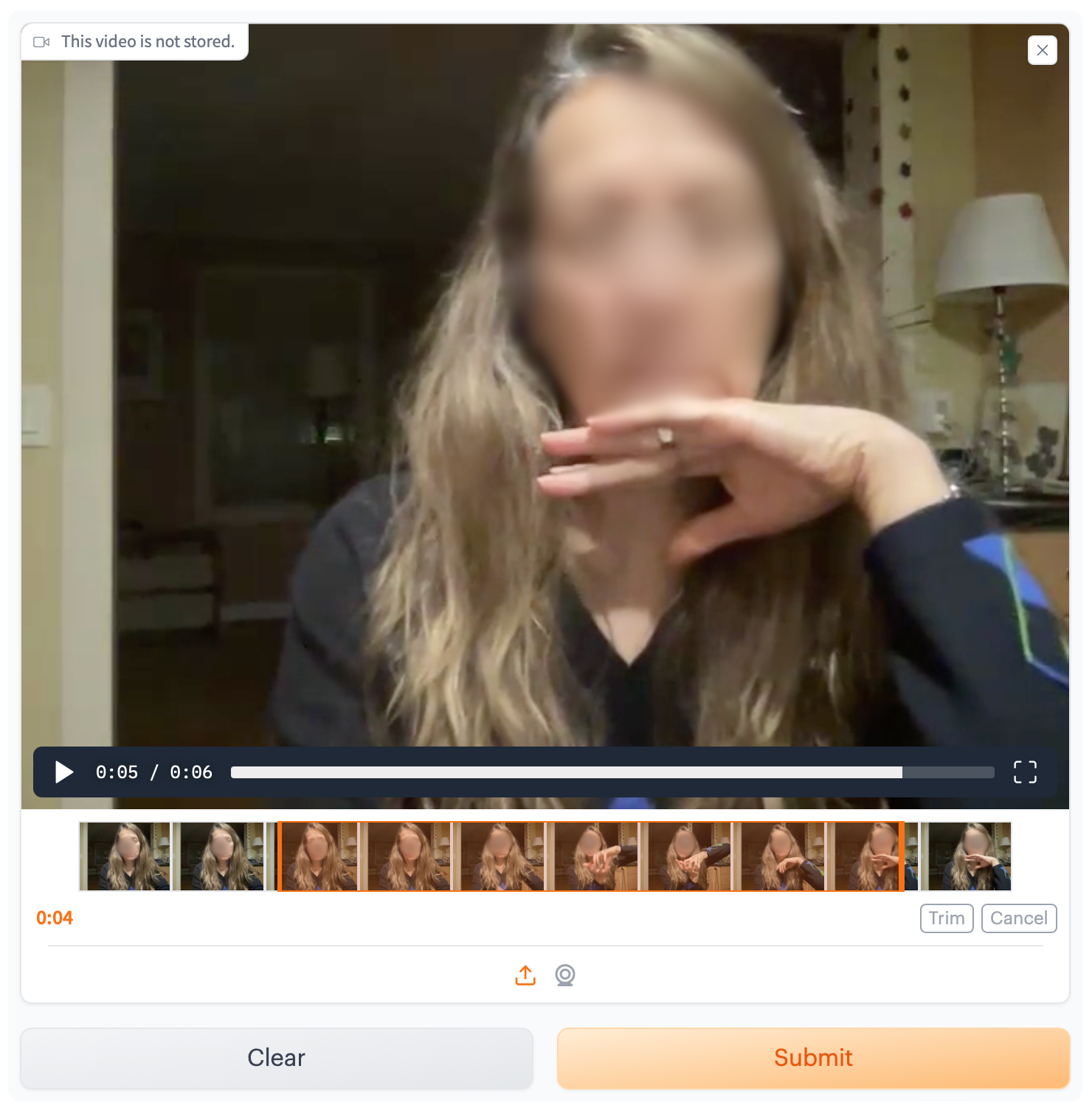}
    \caption{The webcam recording and file upload interface includes built-in tools for video clipping and editing.}
    \Description[Video clipping and editing interface]{The video uploading and recording interface is composed of a preview box and two buttons below the box. A blurred-out frame of a person signing is shown in the box at the top. At the bottom of this box is a progress bar indicating the position of the frame shown in the box. Below this progress bar is a trimming interface with an adjustable start and end position. On the bottom right of this box is a loop and a scissors icon, representing the buttons to reset the trimming or initiate the trimming, respectively. On the bottom center of this box is an arrow pointing up and a webcam icon representing the data source of the uploaded video. Below this box are two buttons: on the left, a gray ‘Clear’ button, and on the right, an orange ‘Submit’ button.}
    \label{fig:videopreview-editing}
\end{figure}

\subsubsection{Video Submission Feedback}\label{feedback}
\label{subsec:video-submission-feedback}

Before the sign recognition AI model analyzes an input video, it is first evaluated for technical and signer visibility. Evaluating technical quality entails verifying that the video resolution and encoding are suitable for processing with the AI model using ffprobe~\cite{ffmpeg}. Manual thresholds were selected for the video size under which the model's performance notably decreased. Evaluating signer visibility entails verifying that the signer is centered in the frame with visible hands and no other people present in the background, as this could confuse the AI model. This is evaluated using MediaPipe~\cite{lugaresi2019mediapipe}, which estimates the pose of people in the input video. Once the poses are estimated, simple rules are applied to determine the number of people visible and the signer's position in the frame. Specifically, a set of conditions verifies that only one person is present, that they are located in the horizontal center of the frame, and that most of their torso, hands, and face are visible (i.e., within the frame and not predicted to be occluded by MediaPipe).



\subsection{User Interface}
\label{design}

We adapted our prototype user interface from \cite{bohacek2023sign} and implemented initial design modifications based on findings from the interviews presented in the same paper.


Our web-based prototype consists of two pages: `Main dictionary interface' and `Detailed results.' On the main page, users can upload or record a video for AI analysis, as shown in Figure~\ref{fig:videopreview-basic} (Appendix~\ref{app:user_interface_figs}). The system first checks for critical issues (e.g., incomplete uploads) and displays an error message if found. Non-critical issues (e.g., video quality, centering, or multiple people) trigger a warning. Section~\ref{subsec:video-submission-feedback} details these checks, with example messages shown in Figure~\ref{fig:errorwarningmessage} (Appendix~\ref{app:user_interface_figs}).

Once uploaded, the recording can be played for user's inspection and trimmed using a simple drag interface if necessary, as shown in Figure~\ref{fig:videopreview-editing}. This feature has been proposed in prior HCI literature on ASL dictionaries \cite{10.1145/3517428.3544883}. The user can then hit the `Clear' button, removing the video, or the `Submit' button, initiating the AI model analysis. When the AI model analysis is initialized, the user is presented with a progress bar, as shown in Figure~\ref{fig:progressbar} (Appendix~\ref{app:user_interface_figs}), indicating the remaining time before a prediction is completed. 

Shown in Figure~\ref{fig:resultsview-main} (Appendix~\ref{app:user_interface_figs}), the compact view presents the top results of the AI analysis, replacing the progress bar once prediction is complete. It shows the top-$7$ signs, with the top-$1$ enlarged and a grid of top-$2$ through top-$7$. Each sign includes the word, example video, textual likelihood, and metadata (movement, number of hands, location, and handshape).

Although the AI model returns likelihoods as percentages, they are converted into confidence labels \cite{google2024pair}. Details on these labels were provided on the landing page\footnote{The language used was: ``\textit{A confidence label is shown next to each prediction, corresponding to the percentual likelihood of the respective sign: 66-100\% to `Probably', 33-66\% to `Possibly', and 0-33\% to Unlikely.}''} at the top of the Main dictionary interface. At the bottom, the `More results' button leads to the Detailed results' page. Shown in Figure~\ref{fig:resultsview-detailed} is the top of the `Detailed analysis' page, presenting the complete results of the AI analysis. Note that only the first few items are presented in this figure; the actual page is scrollable and presents many following signs as well. Each sign item in this view, once again, presents the corresponding word, an example video, textual likelihood, and metadata. Example sign items of varying likelihood are shown in Figure~\ref{fig:resultsview-detailed}. On this page, however, the signs can be filtered by the metadata using the drop-down menus shown at the very top of Figure~\ref{fig:resultsview-detailed}.



\section{Observational Study Method}\label{methods}

\subsection{Study Materials and Design}\label{materials}

A senior ASL educator, who is part of the team, scripted and recorded four videos, each approximately one minute long. The videos featured narratives related to food, countries, and animals, similar to what students would encounter in their ASL II course. The English translations of the narratives are provided in Appendix~\ref{app:narratives}. Our team member also created questions of increasing difficulty and shortlisted three questions per video, totaling 12 questions. These questions were of the open written field type. 

Participants signed a consent form approved by an Institutional Review Board (IRB) before joining the study, and demographic information was collected through a separate form.  To ensure sufficient time for interviews during each study session, participants were asked to watch 2 videos and answer 6 corresponding questions. We used a Latin square protocol to vary the sequence of videos across our 12 participants. They were encouraged to use the dictionary tool to assist in answering the questions, even if they already knew the relevant sign(s). They could use the dictionary multiple times during watching the video or while answering each question. 

After answering each question on the response form, participants were also asked to indicate the extent to which they agreed with the following two statements:

\begin{enumerate}
    \item \emph{I am satisfied with way results are ranked.}: This 5-point Likert scale question (ranging from ``Strongly Disagree'' to ``Strongly Agree'') has been adapted from information retrieval \cite{10.1145/1277741.1277902} and applied to ASL dictionaries research  \cite{10.1145/3491102.3501986}.

    \item \emph{I am satisfied with the search experience.} This 5-point Likert scale question with the same scale was also adapted from prior research on search systems \cite{10.1145/1277741.1277839}.

\end{enumerate}

A semi-structured interview was conducted afterward (details provided in Section \ref{interview-method}). The studies were conducted both in person (n=7) and remotely (n=5). In both formats, participants were asked to share their screens, and the meetings were recorded. A researcher also took notes during the study, documenting participants' usage patterns with the video-based dictionary.

\subsection{Interview Guide}\label{interview-method}

We developed our semi-structured interview guide with feedback from team members specializing in HCI, AI, and ASL pedagogy. Three pilot interviews were conducted to refine the questions. The final interviews consisted of three parts: (1) gathering feedback on their experience using the video-based dictionary for a video-comprehension task, (2) discussing the recording interface, feedback on recordings, and any adjustments made when reattempting a recording, and (3) seeking feedback on the prototype’s depiction of latency, result presentation, confidence display, and suggestions for improving sign search.


After the three main parts, the researcher asked follow-up questions based on notes taken during the interview. These questions were intended to clarify the reasoning behind certain observed behaviors, such as recording videos, navigating results, or answering questions. Participants were also encouraged to share any additional thoughts at the end. Each session was scheduled for 75 minutes, and the average session duration was approximately the same. Participants were paid \$40 for participating in the study.

\subsection{Recruitment and Participants}

All of our participants were students in an ASL II class at a university. The ASL II course at the university where we conducted the study is the second in a sequence of three courses. It focuses on expanded vocabulary, an intermediate-level understanding of grammatical structures, and the continued development of communicative skills. Students are expected to have a foundational knowledge of ASL, basic receptive and expressive signing skills, and an understanding of applied grammar. Our participants included 1 male and 11 females. Two students had a d/Deaf or hard of hearing family member, friend, or relative. The average age of our participants was 20.8 ($\sigma$ = 0.98). The age distribution was as follows: 19 years (2 participants), 20 years (3 participants), 21 years (4 participants), and 22 years (3 participants). On average, our participants had been learning ASL for 3 years ($\sigma$ = 2.730) and took their first course 3 years ($\sigma$ = 2.402) years ago. Six participants mentioned using online sign language dictionaries and three mentioned YouTube videos. Participants reported that, on average, they spend about 4.3 hours ($\sigma$ = 1.494) learning ASL per week. 



\subsection{Qualitative Data Analysis}




Our qualitative data consisted of observations made by the researcher during the study, along with post-interview transcripts, which were both copied into individual Google Docs for all twelve interviews. We employed a reflexive thematic analysis approach \cite{braun2023doing} for data analysis. Initially, two HCI researchers coded the transcripts using comments in Google Docs. We created a codebook following \cite{knoch2020}. The research team then collaboratively discussed the codes in five hour-long meetings, identifying initial themes and refining them into the final set. When presenting our findings, we prioritized user engagement aspects that could not be captured in previous WoZ studies, such as participants' reattempts, rerecordings, feedback on latency, etc. We also remained open to broader insights, including any new benefits or challenges relevant to deployment, such as biases and privacy concerns.

\section{ Findings}\label{findings-qualitative}

\begin{figure}[h!]
    \centering
    \includegraphics[width=0.95\linewidth]{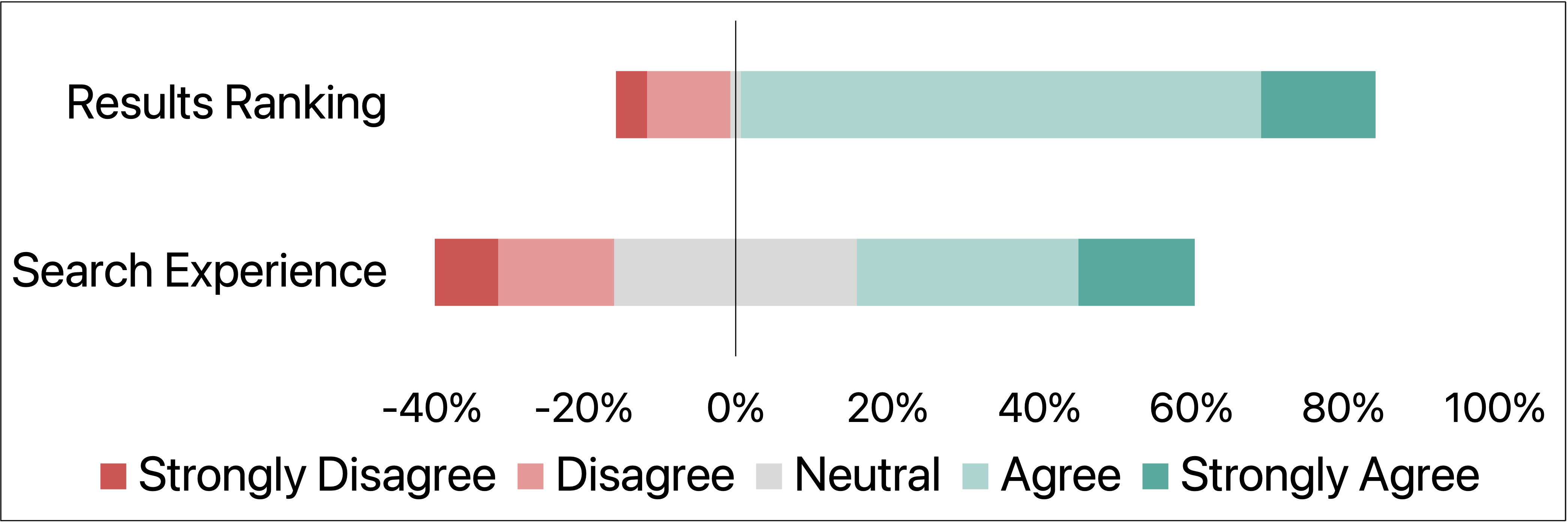}
    \caption{Divergent stacked bar chart summarizing participants' responses to the Likert-scale questions. The responses are centered at 0\%, with the ``Neutral'' category equally distributed between the positive and negative sides of the scale.}
    \Description{
    This stacked horizontal bar chart presents the 'Results Ranking' (top) and 'Search Experience' (bottom) participant responses. Each bar is segmented into the following responses: 'Strongly Disagree' (dark red), 'Disagree' (light red), 'Neutral' (gray), 'Agree' (light green), and 'Strongly Agree' (dark green). The scales range from -40\% to 100\%. A vertical reference line at 0\% divides negative and positive responses.
    }
    \label{divergent-stacked-bar-chart}
\end{figure}

Responses to the two Likert-scale questions were analyzed using summary statistics and visualized using divergent stacked bar charts presented in Figure \ref{divergent-stacked-bar-chart}. 

Our qualitative findings are organized into six themes. The first two themes address \emph{initial positive impressions} and \emph{benefits}. The next three themes focus on users' impressions of different prototype components, including \emph{recording}, \emph{viewing output}, and \emph{performance communication}. The final theme encapsulates findings that extend \emph{beyond the prototype's functional aspects}.

\subsection{Perceived Novelty and Positive Impressions of the Video-Based Dictionary Concept}
Most participants encountered a video-based ASL dictionary for the first time and appreciated its necessity. P5 shared their enthusiasm saying that they \emph{really love the concept} and found it \emph{incredibly helpful and necessary}. They further described the novelty of this experience: ``\textit{I never like using the other dictionaries... I did not know that we're on the way to, like, get the right signs that I want.}'' P9 also appreciated the concept and highlighted how it was useful for them in the comprehension assignment saying ``\textit{there was no way I would have found [signs for] GERMANY or RUSSIA.}'' P2 also noted that they \emph{don't know of any other dictionaries that allow you to search in that manner.}

Participants mentioned being impressed even when the correct result was not ranked first. P8 noted: ``\textit{[When] I was signing slightly different... and it wouldn't pick it up as the first choice. It would put something else... because my hands were bent, I thought it was signing SALAD... it was still by my face... TIGER was one of the... next options. I really liked it.}'' Participants were quickly willing to work with the unpredictable nature of the dictionary, even when the results did not appear at the top of the results list.


\subsection{Easier Search and Learning Benefits}

\subsubsection{Alleviating Challenges with Textually Describing Signs}

Our participants were quick to point out how our tool could alleviate the challenges associated with using text-based queries to describe signs in traditional search engines, which often is unsuccessful. P6 mentioned how they can ``\textit{type in a search engine... but [it] can't ever really translate it back in the same way that this tool lets us do that.}'' P7 compared it to a reverse image search saying:
``\textit{in principle that's cool and useful... it's kind of like reverse image search on Google.}''


\subsubsection{Learning Beyond Sign Look-up}

During the assignment, participants used the tool in several ways that extended beyond just looking up unfamiliar signs. Several participants used the tool to confirm their understanding of signs. P11 mentioned their experience of using the dictionary to confirm their question responses: ``\textit{I kind of did it as a reassurance to see. I was trying to make sure that was the correct thing.}'' 

Our participants also appreciated the opportunity to see multiple examples of people signing, which enriched their learning experience by exposing them to a variety of signing styles. P1 expressed their feedback saying ``\textit{I like how there's multiple different people signing because sometimes if it's the same person, you don't really get to see a variety of it.}''

Similarly, the option to view alternative signs alongside the primary result was well-received. P8 valued this feature for its educational potential, as it allowed them to recognize and understand the subtle differences between the performance of similar signs: ``\textit{...There's like a primary result... oh, but if this isn't right there, here's other options. There's so much like potential for even just seeing like other signs that are similar.}'' They mentioned that they compared similar signs in the results, which motivated them to learn the difference so they can sign it more accurately in the future: ``\textit{But also even just recognizing... I didn't realize how similar these two signs were to each other. I'm gonna be careful when I sign with somebody to make sure that I'm being clear between the difference.}''

 Our findings suggest that participants benefited from the tool in ways beyond easier searching, highlighting its potential to help learners improve skills such as differentiating between signs that share similar linguistic properties.




\subsection{Fine-tuning and Refining Recordings}\label{reattempting}

\subsubsection{Challenge of Recording an Unknown Sign from Memory}

Some of our participants found it challenging to record unfamiliar signs compared to those they already knew. They reported difficulties when trying to mimic video segments in the assignment that they thought corresponded to an unfamiliar sign. P1 described this challenge: ``\textit{The sign for GERMANY, I hadn't seen that before. It took a lot longer to record video... I wasn't signing it correctly the first few times. And I noticed that it would recognize that... my hands were sprawled out, so it thought I was doing TIGER.}''

We observed that at least 8 participants found it challenging to isolate signs within segments of unfamiliar videos. Four participants used the trimming feature in our video submission interface to attempt searches with different sub-segments of the videos they recorded. We also noticed that some participants slowed down their performance of the signs when they did not find the correct result on the first page. These findings highlight the challenges of video-based querying, particularly when users are unable to accurately perform a sign from memory, despite the shorter querying process.

\subsubsection{Adjusting Camera and Body}

During the study, participants adjusted their camera, body, or both, especially after receiving feedback. In post-interviews, they described strategies for optimizing recordings, focusing on hand configuration, sign position, movement, and facial expressions.

P1 described their experience of repositioning their camera and signing slowly: ``\textit{I tried to reposition the camera, so it got more of my body and less of my face if I was standing a bit lower. And I would also try to move my hands a little slower, so... the video would be able to pick up on my hands better.}''

Participants employed different strategies depending on the sign. P8 explained how they adjusted their distance from the camera to open up the ``signing space'':

\begin{quote}
    `\textit{[If] sign was on the face, I would just do... But if I had something on the body, then I would try to move back and... adjust my camera and... field of view... It would come up just maybe not as the primary... still one of the first few options... I would [also] move... further back to open my signing space to the camera.}''
\end{quote}

In contrast, P5 spoke about getting closer to the camera for certain signs but was unsure if it was useful: ``\textit{I gotta get close to the camera... definitely adjust a little bit at the beginning... trying to make sure they see hands. I don't know if that helps or hurt it.}''

During the study, participants were free to move the laptop to their desired location, exposing the camera to different backgrounds. Some participants realized that the background could affect the quality of the dictionary output, especially when the prototype provided feedback on this. P4 noted: ``\textit{If things didn’t come up, or it didn’t recognize me, then I realized I needed a very clean background. I definitely improved once I shifted over to have this as my background.}''

Our findings show that participants were willing to refine their input to improve the dictionary's performance. We also found that at least some participants quickly understood the need to adjust their position relative to the camera based on the linguistic properties of a sign, such as its use of space.

\subsubsection{Reattempting Recording}

 Participants shared their reattempting experiences: ``\textit{probably about three times before I gave up?}'' (P1), ``\textit{Um, probably two, three.}'' (P8), and ``\textit{Um, feel like three or four depends... if it's not understanding me? Or I'm doing wrong?}'' (P10).Some of the participants mentioned that they can be more persistent if they see that the results are changing: ``\textit{If I tried three times, and it gave me like the exact same results every single time, then I'd probably stop. But if their results were changing, and it was giving me different signs, then I'd probably [try] five different times.}''

For some participants, the decision to continue trying was also based on the quality of the output, as noted by P4: ``\textit{I spent a really long time trying to figure out. So I would probably keep trying... if I'm sure I'm doing the right hand movements and everything and not getting a clearer picture, I'd probably give up.}'' The context also determined how many attempts they made, with P9 noting potential differences between exams and homework settings: ``\textit{If I'm doing it on the exam, I'll think over it, keep doing it. But if I'm doing it for homework, I'd give it four or five [tries].}'' P10 suggested the need for more specific feedback on the recorded submission: ``\textit{maybe something about your performance of sign... I did bird, they're saying, I said, Greece. So then how is it that? How could they correct me.}''

The findings suggest that participants are willing to reattempt recordings, with their willingness potentially influenced by the dictionary's output and the importance of the task they are trying to accomplish. Additionally, our results indicate that providing more feedback on the accuracy of their performance, based on context from the assignment or user input, could be beneficial.



\subsection{Navigating Unpredictability of Output}\label{unpredictability}

Participants, new to video-based ASL dictionaries, were surprised by both successful and failed searches. This theme reflects feedback on the AI dictionary’s non-deterministic and unpredictable nature.

\subsubsection{Dealing with Inconsistent Dictionary Performance}

Some of our participants were surprised by the tool's inconsistent performance. For example, P2 said: ``\textit{It seemed to work really well for some signs and then not at all for other signs.... Not sure if I wasn't doing it right... or not clear enough video. But for sometimes it came right up.}''

While some participants like P2 did not blame a specific entity for a failed search, about half of the participants blamed themselves when they could not find the sign. For example, P3 attributed unsuccessful searches to their understanding of the video: ``\textit{I think that any of the issues are more so like my comprehension of the video and not understanding certain signs, but not necessarily an indication that... the tool itself was unhelpful.}''

About half of the participants were able to make more informed decisions about whether to attribute unsuccessful searches to the prototype or their recording, based on the feedback on video-input confidence levels in results.

\subsubsection{Presentation of Confidence Levels}

To help users manage frustration when they cannot find the correct sign, we had added AI's outputted confidence with the output in the results. P2 appreciated the clarity provided by probability indicators: ``\textit{I liked that it says like, it gives you like the probably are not likely so you kind of know how sure it is instead of just like listing that without that."}

At least two participants highlighted that these confidence values could be confusing when not appropriately assigned, especially for novice learners: ``\textit{When TIGER was number one, and it had that but had unlikely as like a possibility... if I didn't know that, like, I probably have been very confusing.}'' A minority of participants, like P9, requested more detailed probability information:
\begin{quote}
    ``\textit{Instead of just saying... `likely' `unlikely' maybe give like a percent of how unlikely or likely just because like, it's pretty vague, like saying, probably like, I know, it's not exact, but maybe like, this is 20\%. Your sign or this is like set if I see like 75[\%] I'm like, okay, that's probably my sign, but like probably is kind of vague.}''
\end{quote}

One of the participants mentioned how presenting more results on the landing page might be more advantageous, especially when the dictionary prototype does not give the right result. P3 suggested making more results immediately visible to avoid the perception that important results are being overlooked: 

\begin{quote}
    ``\textit{..probably be easier to not click on a different thing to load more results. [Even] If wouldn't sign the exact same thing... could be viewed as the same as Google search—if you don't find something on the first page, 90\% of people won't even bother going to the second... }''
\end{quote}

We applied best research-based industry practices in presenting confidence levels and splitting results between a landing page and a ``detailed analysis'' page for further results. While the majority of participants appreciated these choices, some suggested alternative approaches which could suggest that there is no one-size-fits-all solution in this context and need for customization.

\subsection{Communicating State and Latency}

Another unique aspect of interaction that our fully functioning prototype afforded was testing for latency. We conveyed feedback on the prototype status and the processing of results using a progress bar. P1 thought that we communicated prototype state and latency well: ``\textit{That's a good amount of information. If it didn't have that loading bar, I'd probably think that the website was glitching out, and would refresh it. So the bar showing how long it's taking was nice.}'' 

Other participants mentioned that even waiting for a few seconds can be annoying. P6 said: ``\textit{I mean, I guess I didn't like waiting six seconds for it to load. But that's just because I'm impatient. But other than that, I thought it was really actually very impressive.}'' 

Our findings suggest that our dictionary effectively communicates its current state, but the latency may be perceived differently by various users.

\subsection{Beyond Functional Concerns}

\subsubsection{Privacy Concerns}

After recording and submitting several videos, and reviewing the privacy considerations, we asked participants about their privacy concerns regarding the dictionary. Most participants, including P8 appeared less concerned about privacy, noting the ease of permission settings on their browsers:
``\textit{All I had to do was like hit allow through Google Chrome.}'' Further elaborating on their comfort with the data handling, they said: ``\textit{I mean, there are disclaimers that videos are kept, but I feel like everything is already out there. So I wouldn't avoid using it... It wouldn't steer me away.}''

Conversely, P9 expressed that they would like to see even more disclaimers about their video being recorded, processed, and stored: ``\textit{Like a button that  told you like your data is being saved.... I'm sure it's not going into some like video that you put on the screen. But like I think that's really important to protect people especially... how controversial it is.}''

The findings suggest that despite reassurances in the application about not storing raw video data, some participants may still have concerns about frequently submitting multiple recordings when searching for unfamiliar signs.

\subsubsection{Recurring Unrelated Signs in Results}

During our observational analysis, an interesting pattern emerged where certain signs frequently reoccurred in the results list for specific participants, seemingly unrelated to their actual submissions. The participants themselves noted this occurrence. For example, P4 observed: ``\textit{I also think that some of the signs that I wasn't trying to sign came up as an example, like multiple times. For instance, FISH came up for four different signs that I was trying to sign. I wasn't looking for the sign for FISH.}'' Similarly, P2 frequently encountered a different sign, potentially indicating some inadvertent biases learned by the model that require deeper evaluation. P2 said: ``\textit{For whatever reason, MILK would always come up as the sign, even if I was doing a sign that didn't really even look like that.}'' This finding indicates that the underlying model might be learning characteristics of the video beyond the sign's performance and could also be encoding biases.




\section{Discussion}\label{discussion}

While previous HCI research has explored optimal interface design for ASL dictionaries using WoZ prototypes \cite{10.1145/3491102.3501986, 10.1145/3517428.3544883}, and AI research has focused on the performance of underlying sign-recognition models \cite{wilcox1994multimedia, camgoz2020sign, elliott2011search, yanovich2016detection, 10.5555/1880751.1880787, camgoz2018neural, pu2018dilated, RASTGOO2021113794, NEURIPS2023_00dada60}, little research has addressed the integration of design recommendations into fully functional prototype or examined how novice ASL learners utilize these dictionaries in an educational context.


WoZ studies show ASL learners prefer video comprehension tools over feature-based baselines \cite{10.1145/3517428.3544883}. Similarly, hybrid search, which combines video-based lookup with feature-based refinement, is also favored by participants in prior work \cite{10.1145/3491102.3501986}. We adopted this approach, focusing on building a functional dictionary prototype that supports ASL learners in real-world use and conducting an observational studyrather than directly comparing it to a baseline \cite{10.1145/3517428.3544883}. This approach provided new insights into recording challenges and the unpredictability of search results. Our ASL dictionary prototype was well received and is being released for research. It can support future HCI studies on dictionary interfaces and aid AI researchers in testing isolated sign language models.



\subsection{Benefits of Video-based ASL Dictionary}\label{benefits}


In our study, participants were positively surprised by the tool’s concept and performance, likening it to novel search systems like Google’s reverse image search. Most were unfamiliar with video-based ASL dictionaries, which represent a major leap in ASL learning \cite{bragg2015user} and enhancing the employability of individuals who use ASL. Our findings confirm prior benefits, especially for comprehension tasks like identifying or verifying signs \cite{10.1145/3517428.3544883, 10.1145/3491102.3501986}, while also revealing new advantages, such as improving sign accuracy and distinguishing similar signs. We hope researchers expand our prototype's use in different educational and professional contexts.

\subsection{Usage of Video-Based Dictionary}\label{usage}

The observational study aimed to understand user interaction with the prototype and improve their experience. The findings revealed both usage patterns (RQ1) and key challenges of the video-based ASL dictionary (RQ2).

\subsubsection{Recording Video Submissions}

As previously noted, a key difference between our study and prior WoZ studies is that our functional prototype does not guarantee that the desired result will appear in the results list \cite{TACCESS, 10.1145/3491102.3501986}. Additionally, although participants had access to raw videos they were attempting to transcribe, unlike in some prior studies, the videos were not presented in isolation \cite{10.1145/3491102.3501986, alonzo2019effect}, nor could users submit a clip of the video as a search query \cite{10.1145/3517428.3544883}. Recording an unknown sign from memory proved challenging for some users, however, most participants persisted with re-submissions and mentioned being comfortable resubmitting at least three times. This was consistent with quantitative findings which revealed that participants conducted 2.32 searches during each continuous recording session.


\subsection{Strategies for Better Video Submissions}  

Participants employed various strategies to improve their recordings. Some participants adjusted their position or repositioned the camera to accommodate different signing spaces. For signs performed on the body, they moved farther from the camera, while for signs near the face, they moved closer to ensure better visibility. Others noticed that background clutter affected video quality, with or without prototype feedback. To address this, they adjusted their laptop placement, leading to improved results. Cluttered backgrounds are a known challenge in AI-based body movement recognition \cite{jegham2020vision}. Our prototype provided feedback only after video submission, but our findings suggest that ASL dictionaries could benefit from real-time overlays during recording to guide users more effectively.  

A few participants slowed their signing to aid comprehension, a common strategy with human interlocutors. However, this was less effective in improving results. Research on ML-based sign fluency feedback \cite{9413126} suggests that future ASL dictionaries could integrate such models to provide real-time performance guidance.  

Our findings highlight the need for clear guidelines to help users record quality videos on personal computers. These should cover distance from the screen, optimal use of space, and automatic background feedback. Guidelines could be shared during onboarding and tailored to different devices.

\subsubsection{Navigating Latency, Results, and Post-Query Refinements}


Our results align with prior web search research, showing that users accustomed to fast systems (like feature-based dictionaries) notice delays more than those using slower systems. Future work should explore delay thresholds \cite{10.1145/2600428.2609627}. Participants appreciated confidence indicators but some preferred exact percentages, despite HAI guidelines advising against them \cite{google2024pair}. Future designs could allow users to toggle between word-based and numerical confidence levels. Some preferred all results on the landing page or a customizable layout. Prior HCI research found that initial result count impacts user experience \cite{TACCESS}, with DCG values similar to our study (0.8). Our findings reinforce that landing page layout affects user satisfaction.

\subsection{Considerations for Deployment}\label{privacy}

Our DCG analysis suggests significant improvements in ASL dictionaries over prior work \cite{athitsos2010large}. Thematic analysis indicates users are willing to use the dictionary at current accuracy levels. 

Interviews also explored non-functional concerns like privacy. Participants raised privacy considerations, particularly regarding our landmark-based AI model, which processes video with MediaPipe and discards it post-analysis, minimizing retention risks. Some preferred explicit recording permissions. Future research could explore overlays and local pose estimation to avoid video transmission, sending only extracted landmarks.

A surprising finding was the presence of unrelated signs in results, suggesting potential biases in the AI model. These biases may stem from the pose estimation model used in preprocessing. Addressing this requires more diverse training data or augmentation techniques \cite{bragg2019sign, 10.1145/3436996}. Since sign language-specific augmentation methods are limited, future work could explore synthetic data generation to improve accuracy and reduce bias. Additionally, less reliable signs could be flagged in results or have enhanced confidence labels to reflect model limitations.

\section{Limitation and Future Work}

We studied the use of our prototype by ASL learners from an ASL II class at the same institution, ensuring their prior experiences were relatively comparable. Future research could broaden participant diversity by including individuals who have not attended formal ASL classes or younger students from high schools to see if these groups would similarly benefit from the proposed designs.

We took notes on user behavior during the study and conducted post-interviews to better understand their experience. Future researchers could enhance the dictionary to directly collect more precise user behavior data, such as fine-tuning submitted videos, trimming them, or scrolling through the results list to reveal additional usage patterns.

Previous WoZ studies have shown that video-based dictionaries are preferred over feature-based ones \cite{10.1145/3517428.3544883}. Thus, we conducted an observational study instead of a baseline comparison, focusing on novel features in our functional prototype. Future research could compare fully functional video-based dictionaries, especially with larger vocabularies, to feature-based alternatives.

Our comprehension tasks targeted three vocabulary sets from an ASL course. Future research could explore its benefits for larger vocabularies and more complex translation tasks. With improved performance enabling effective interactions, longitudinal studies across course modules could reveal user patterns and challenges that emerge over extended use.

Automated sign language recognition has historically focused on ASL, with most methods, datasets, and benchmarks dedicated to it. While other sign languages have gained attention, a performance gap remains. Future AI and computer vision research should work to close this gap. HCI research should also explore integrating non-ASL models into education and assess whether performance disparities hinder dictionary deployment for these languages.



\section{Conclusion}\label{conclusion}

Prior work has explored the design of isolated sign-recognition systems and optimal design parameters for ASL dictionaries separately. However, there is limited research examining the benefits and usage of a fully automated video-based ASL dictionary for learners engaged in comprehension tasks. To fill this gap, we conducted an observational study with 12 ASL learners using a dictionary built on state-of-the-art sign-recognition technology and informed by prior HCI research. We release our ASL dictionary prototype with this publication. Our findings highlight diverse usage patterns, including sign recording, result retrieval, and perceptions of system latency, while also revealing some outstanding concerns such as potential bias and privacy issues. These insights guide future design and effective deployment of ASL dictionaries in both educational and professional settings.


\bibliographystyle{plain}
\bibliography{sample-base}

\begin{thebibliography}{10}

\bibitem{10.1145/1277741.1277902}
Azzah Al-Maskari, Mark Sanderson, and Paul Clough.
\newblock The relationship between ir effectiveness measures and user satisfaction.
\newblock In {\em Proceedings of the 30th Annual International ACM SIGIR Conference on Research and Development in Information Retrieval}, SIGIR '07, page 773–774, New York, NY, USA, 2007. Association for Computing Machinery.

\bibitem{alonzo2019effect}
Oliver Alonzo, Abraham Glasser, and Matt Huenerfauth.
\newblock Effect of automatic sign recognition performance on the usability of video-based search interfaces for sign language dictionaries.
\newblock In {\em The 21st International ACM SIGACCESS Conference on Computers and Accessibility}, ASSETS '19, page 56–67, New York, NY, USA, 2019. Association for Computing Machinery.

\bibitem{10.1145/3290605.3300233}
Saleema Amershi, Dan Weld, Mihaela Vorvoreanu, Adam Fourney, Besmira Nushi, Penny Collisson, Jina Suh, Shamsi Iqbal, Paul~N. Bennett, Kori Inkpen, Jaime Teevan, Ruth Kikin-Gil, and Eric Horvitz.
\newblock Guidelines for human-ai interaction.
\newblock In {\em Proceedings of the 2019 CHI Conference on Human Factors in Computing Systems}, CHI '19, page 1–13, New York, NY, USA, 2019. Association for Computing Machinery.

\bibitem{10.1145/2600428.2609627}
Ioannis Arapakis, Xiao Bai, and B.~Barla Cambazoglu.
\newblock Impact of response latency on user behavior in web search.
\newblock In {\em Proceedings of the 37th International ACM SIGIR Conference on Research \& Development in Information Retrieval}, SIGIR '14, page 103–112, New York, NY, USA, 2014. Association for Computing Machinery.

\bibitem{athitsos2010large}
Vassilis Athitsos, Carol Neidle, Stan Sclaroff, Joan Nash, Alexandra Stefan, Ashwin Thangali, Haijing Wang, and Quan Yuan.
\newblock Large lexicon project: American sign language video corpus and sign language indexing/retrieval algorithms.
\newblock In {\em Workshop on the Representation and Processing of Sign Languages: Corpora and Sign Language Technologies (CSLT)}, volume~2, pages 11--14, Valletta, Malta, 2010. European Language Resources Association (ELRA).

\bibitem{bohacek2023sign}
Matyas Bohacek and Saad Hassan.
\newblock Sign spotter: Design and initial evaluation of an automatic video-based american sign language dictionary system.
\newblock In {\em Proceedings of the 25th International ACM SIGACCESS Conference on Computers and Accessibility}, pages 1--5, 2023.

\bibitem{bohavcek2022sign}
Maty{\'a}{\v{s}} Boh{\'a}{\v{c}}ek and Marek Hr{\'u}z.
\newblock Sign pose-based transformer for word-level sign language recognition.
\newblock In {\em Proceedings of the IEEE/CVF winter conference on applications of computer vision}, pages 182--191, 2022.

\bibitem{bohacek2022sign}
Maty{\'a}{\v{s}} Boh{\'a}{\v{c}}ek and Marek Hr{\'u}z.
\newblock Sign pose-based transformer for word-level sign language recognition.
\newblock In {\em Proceedings of the IEEE/CVF winter conference on applications of computer vision}, pages 182--191, 2022.

\bibitem{10.1145/3436996}
Danielle Bragg, Naomi Caselli, Julie~A. Hochgesang, Matt Huenerfauth, Leah Katz-Hernandez, Oscar Koller, Raja Kushalnagar, Christian Vogler, and Richard~E. Ladner.
\newblock The fate landscape of sign language ai datasets: An interdisciplinary perspective.
\newblock {\em ACM Trans. Access. Comput.}, 14(2), jul 2021.

\bibitem{bragg2019sign}
Danielle Bragg, Oscar Koller, Mary Bellard, Larwan Berke, Patrick Boudreault, Annelies Braffort, Naomi Caselli, Matt Huenerfauth, Hernisa Kacorri, Tessa Verhoef, Christian Vogler, and Meredith Ringel~Morris.
\newblock Sign language recognition, generation, and translation: An interdisciplinary perspective.
\newblock In {\em The 21st International ACM SIGACCESS Conference on Computers and Accessibility}, ASSETS '19, page 16–31, New York, NY, USA, 2019. Association for Computing Machinery.

\bibitem{bragg2015user}
Danielle Bragg, Kyle Rector, and Richard~E. Ladner.
\newblock A user-powered american sign language dictionary.
\newblock In {\em Proceedings of the 18th ACM Conference on Computer Supported Cooperative Work and Social Computing}, CSCW '15, page 1837–1848, New York, NY, USA, 2015. Association for Computing Machinery.

\bibitem{braun2023doing}
Virginia Braun, Victoria Clarke, Nikki Hayfield, Louise Davey, and Elizabeth Jenkinson.
\newblock Doing reflexive thematic analysis.
\newblock In {\em Supporting research in counselling and psychotherapy: Qualitative, quantitative, and mixed methods research}, pages 19--38. Springer, 2023.

\bibitem{components}
Fabian Bross.
\newblock Chereme. in: Hall, t. a. pompino-marschall, b. (ed.): Dictionaries of linguistics and communication science (wörterbücher zur sprach- und kommunikationswissenschaft, wsk). volume: Phonetics and phonology. berlin, new york: Mouton de gruyter.
\newblock {\em Proceedings of the XVIII EURALEX International Congress}, 1(1):9, 01 2015.

\bibitem{10.1145/1229390.1229401}
Fabio Buttussi, Luca Chittaro, and Marco Coppo.
\newblock Using web3d technologies for visualization and search of signs in an international sign language dictionary.
\newblock In {\em Proceedings of the Twelfth International Conference on 3D Web Technology}, Web3D '07, page 61–70, New York, NY, USA, 2007. Association for Computing Machinery.

\bibitem{camgoz2018neural}
Necati~Cihan Camg{\"{o}}z, Simon Hadfield, Oscar Koller, Hermann Ney, and Richard Bowden.
\newblock Neural sign language translation.
\newblock In {\em 2018 {IEEE} Conference on Computer Vision and Pattern Recognition, {CVPR} 2018, Salt Lake City, UT, USA, June 18-22, 2018}, pages 7784--7793, New York, New York, US, 2018. {IEEE} Computer Society.

\bibitem{camgoz2020sign}
Necati~Cihan Camg{\"{o}}z, Oscar Koller, Simon Hadfield, and Richard Bowden.
\newblock Sign language transformers: Joint end-to-end sign language recognition and translation.
\newblock In {\em 2020 {IEEE/CVF} Conference on Computer Vision and Pattern Recognition, {CVPR} 2020, Seattle, WA, USA, June 13-19, 2020}, pages 10020--10030, New York, New York, US, 2020. {IEEE}.

\bibitem{10.1145/3391613}
Jieshan Chen, Chunyang Chen, Zhenchang Xing, Xin Xia, Liming Zhu, John Grundy, and Jinshui Wang.
\newblock Wireframe-based ui design search through image autoencoder.
\newblock {\em ACM Trans. Softw. Eng. Methodol.}, 29(3), jun 2020.

\bibitem{daly2021user}
Elizabeth~M Daly, Massimiliano Mattetti, {\"O}znur Alkan, and Rahul Nair.
\newblock User driven model adjustment via boolean rule explanations.
\newblock In {\em Proceedings of the AAAI Conference on Artificial Intelligence}, volume~35, pages 5896--5904, 2021.

\bibitem{dennis2023ai}
Alan~R Dennis, Akshat Lakhiwal, and Agrim Sachdeva.
\newblock Ai agents as team members: Effects on satisfaction, conflict, trustworthiness, and willingness to work with.
\newblock {\em Journal of Management Information Systems}, 40(2):307--337, 2023.

\bibitem{desai2024asl}
Aashaka Desai, Lauren Berger, Fyodor Minakov, Nessa Milano, Chinmay Singh, Kriston Pumphrey, Richard Ladner, Hal Daum{\'e}~III, Alex~X Lu, Naomi Caselli, et~al.
\newblock Asl citizen: A community-sourced dataset for advancing isolated sign language recognition.
\newblock {\em Advances in Neural Information Processing Systems}, 36, 2024.

\bibitem{ffmpeg}
FFmpeg Developers.
\newblock Ffmpeg.
\newblock \url{https://ffmpeg.org/}, 2024.
\newblock Version 6.0.

\bibitem{dhole2024queryexplorer}
Kaustubh~D Dhole, Shivam Bajaj, Ramraj Chandradevan, and Eugene Agichtein.
\newblock Queryexplorer: An interactive query generation assistant for search and exploration.
\newblock {\em arXiv preprint arXiv:2403.15667}, 2024.

\bibitem{elliott2011search}
Ralph Elliott, Helen Cooper, John Glauert, Richard Bowden, and Fran\c{c}ois Lefebvre-Albaret.
\newblock Search-by-example in multilingual sign language databases.
\newblock In {\em Proceedings of the Second International Workshop on Sign Language Translation and Avatar Technology (SLTAT)}, Dundee, Scotland, October 23 2011. SLTAT.

\bibitem{NCED}
National~Center for Education Statistics~(NCES).
\newblock Digest of education statistics number and percentage distribution of course enrollments in languages other than english at degree-granting postsecondary institutions, by language and enrollment level: Selected years, 2002 through 2016, 2018.

\bibitem{fuertes2006bilingual}
Jos{\'e}~L. Fuertes, {\'A}ngel~L. Gonz{\'a}lez, Gonzalo Mariscal, and Carlos Ruiz.
\newblock Bilingual sign language dictionary.
\newblock In Klaus Miesenberger, Joachim Klaus, Wolfgang~L. Zagler, and Arthur~I. Karshmer, editors, {\em Computers Helping People with Special Needs}, pages 599--606, Berlin, Heidelberg, 2006. Springer Berlin Heidelberg.

\bibitem{goldberg2015enrollments}
David Goldberg, Dennis Looney, and Natalia Lusin.
\newblock Enrollments in languages other than english in united states institutions of higher education, fall 2013., 2015.

\bibitem{google2024pair}
{Google PAIR}.
\newblock Google pair: People + ai research, 2024.
\newblock Accessed: 2024-09-08.

\bibitem{hall2017language}
Wyatte~C Hall, Leonard~L Levin, and Melissa~L Anderson.
\newblock Language deprivation syndrome: A possible neurodevelopmental disorder with sociocultural origins.
\newblock {\em Social psychiatry and psychiatric epidemiology}, 52(6):761--776, 2017.

\bibitem{hameed2022privacy}
Hira Hameed, Muhammad Usman, Muhammad~Zakir Khan, Amir Hussain, Hasan Abbas, Muhammad~Ali Imran, and Qammer~H Abbasi.
\newblock Privacy-preserving british sign language recognition using deep learning.
\newblock In {\em 2022 44th Annual International Conference of the IEEE Engineering in Medicine \& Biology Society (EMBC)}, pages 4316--4319. IEEE, 2022.

\bibitem{hassan2020effect}
Saad Hassan, Oliver Alonzo, Abraham Glasser, and Matt Huenerfauth.
\newblock Effect of ranking and precision of results on users’ satisfaction with search-by-video sign-language dictionaries.
\newblock In {\em Sign Language Recognition, Translation and Production (SLRTP) Workshop-Extended Abstracts}, volume~4, Virtual, 2020. Computer Vision -- ECCV 2020 Workshops.

\bibitem{TACCESS}
Saad Hassan, Oliver Alonzo, Abraham Glasser, and Matt Huenerfauth.
\newblock Effect of sign-recognition performance on the usability of sign-language dictionary search.
\newblock {\em ACM Trans. Access. Comput.}, 14(4), oct 2021.

\bibitem{10.1145/3470650}
Saad Hassan, Oliver Alonzo, Abraham Glasser, and Matt Huenerfauth.
\newblock Effect of sign-recognition performance on the usability of sign-language dictionary search.
\newblock {\em ACM Trans. Access. Comput.}, 14(4), oct 2021.

\bibitem{10.1145/3517428.3544883}
Saad Hassan, Akhter~Al Amin, Calu\~{a} de~Lacerda~Pataca, Diego Navarro, Alexis Gordon, Sooyeon Lee, and Matt Huenerfauth.
\newblock Support in the moment: Benefits and use of video-span selection and search for sign-language video comprehension among asl learners.
\newblock In {\em Proceedings of the 24th International ACM SIGACCESS Conference on Computers and Accessibility}, ASSETS '22, New York, NY, USA, 2022. Association for Computing Machinery.

\bibitem{10.1145/3491102.3501986}
Saad Hassan, Akhter~Al Amin, Alexis Gordon, Sooyeon Lee, and Matt Huenerfauth.
\newblock Design and evaluation of hybrid search for american sign language to english dictionaries: Making the most of imperfect sign recognition.
\newblock In {\em Proceedings of the 2022 CHI Conference on Human Factors in Computing Systems}, CHI '22, New York, NY, USA, 2022. Association for Computing Machinery.

\bibitem{hoffmeister2000piece}
Robert~J Hoffmeister.
\newblock {\em A piece of the puzzle: ASL and reading comprehension in deaf children}.
\newblock Mahwah, N.J. : Lawrence Erlbaum Associates, New Jersey, USA, 2000.

\bibitem{huenerfauth2017evaluation}
Matt Huenerfauth, Elaine Gale, Brian Penly, Sree Pillutla, Mackenzie Willard, and Dhananjai Hariharan.
\newblock Evaluation of language feedback methods for student videos of american sign language.
\newblock {\em ACM Transactions on Accessible Computing (TACCESS)}, 10(1):1--30, 2017.

\bibitem{10.1145/1277741.1277839}
Scott~B. Huffman and Michael Hochster.
\newblock How well does result relevance predict session satisfaction?
\newblock In {\em Proceedings of the 30th Annual International ACM SIGIR Conference on Research and Development in Information Retrieval}, SIGIR '07, page 567–574, New York, NY, USA, 2007. Association for Computing Machinery.

\bibitem{10.5555/1880751.1880787}
Kabil Jaballah and Mohamed Jemni.
\newblock Toward automatic sign language recognition from web3d based scenes.
\newblock In {\em Proceedings of the 12th International Conference on Computers Helping People with Special Needs}, ICCHP'10, page 205–212, Berlin, Heidelberg, 2010. Springer-Verlag.

\bibitem{jegham2020vision}
Imen Jegham, Anouar~Ben Khalifa, Ihsen Alouani, and Mohamed~Ali Mahjoub.
\newblock Vision-based human action recognition: An overview and real world challenges.
\newblock {\em Forensic Science International: Digital Investigation}, 32:200901, 2020.

\bibitem{knoch2020}
Carsten Knoch.
\newblock Qualitative data analysis with microsoft word, comments, and python (updated), 2020.
\newblock Accessed: 2024-09-12.

\bibitem{lapiak}
J.~Lapiak.
\newblock Handspeak, 2021.

\bibitem{lugaresi2019mediapipe}
Camillo Lugaresi, Jiuqiang Tang, Hadon Nash, Chris McClanahan, Esha Uboweja, Michael Hays, Fan Zhang, Chuo-Ling Chang, Ming~Guang Yong, Juhyun Lee, et~al.
\newblock Mediapipe: A framework for building perception pipelines.
\newblock {\em arXiv preprint arXiv:1906.08172}, 2019.

\bibitem{10.1145/3491102.3517565}
Amama Mahmood, Jeanie~W Fung, Isabel Won, and Chien-Ming Huang.
\newblock Owning mistakes sincerely: Strategies for mitigating ai errors.
\newblock In {\em Proceedings of the 2022 CHI Conference on Human Factors in Computing Systems}, CHI '22, New York, NY, USA, 2022. Association for Computing Machinery.

\bibitem{majetic2017proposing}
Klara Majeti{\'c} and Petra Bago.
\newblock Proposing an instrument for evaluation of online dictionaries of sign languages.
\newblock {\em INTEGRATING ICTIN SOCIETY}, 1(1):189, 2017.

\bibitem{mccarty2004notation}
Amy~L McCarty.
\newblock Notation systems for reading and writing sign langusage.
\newblock {\em The Analysis of verbal behavior}, 20:129--134, 2004.

\bibitem{mckee2015emergency}
Michael~M McKee, Paul~C Winters, Ananda Sen, Philip Zazove, and Kevin Fiscella.
\newblock Emergency department utilization among deaf american sign language users.
\newblock {\em Disability and Health Journal}, 8(4):573--578, 2015.

\bibitem{sign_bsl}
Daniel Mitchell.
\newblock British sign language bsl dictionary, 2021.

\bibitem{Neidle_2012_Challenges}
C.~Neidle, Ashwin Thangali, and Stan Sclaroff.
\newblock Challenges in development of the american sign language lexicon video dataset ({ASLLVD}) corpus.
\newblock 2012.

\bibitem{Paszke2019PyTorchAI}
Adam Paszke, Sam Gross, Francisco Massa, Adam Lerer, James Bradbury, Gregory Chanan, Trevor Killeen, Zeming Lin, Natalia Gimelshein, Luca Antiga, Alban Desmaison, Andreas K{\"o}pf, Edward Yang, Zach DeVito, Martin Raison, Alykhan Tejani, Sasank Chilamkurthy, Benoit Steiner, Lu~Fang, Junjie Bai, and Soumith Chintala.
\newblock Pytorch: An imperative style, high-performance deep learning library.
\newblock {\em ArXiv}, abs/1912.01703, 2019.

\bibitem{paudyal2019learn2sign}
Prajwal Paudyal, Junghyo Lee, Azamat Kamzin, Mohamad Soudki, Ayan Banerjee, and Sandeep~KS Gupta.
\newblock Learn2sign: Explainable ai for sign language learning.
\newblock In {\em IUI Workshops}, 2019.

\bibitem{pu2018dilated}
Junfu Pu, Wengang Zhou, and Houqiang Li.
\newblock Dilated convolutional network with iterative optimization for continuous sign language recognition.
\newblock In {\em Proceedings of the Twenty-Seventh International Joint Conference on Artificial Intelligence, {IJCAI-18}}, pages 885--891, Stockholm, Sweden, 7 2018. International Joint Conferences on Artificial Intelligence Organization.

\bibitem{quinto2011teaching}
David Quinto-Pozos.
\newblock Teaching american sign language to hearing adult learners.
\newblock {\em Annual Review of Applied Linguistics}, 31:137--158, 2011.

\bibitem{RASTGOO2021113794}
Razieh Rastgoo, Kourosh Kiani, and Sergio Escalera.
\newblock Sign language recognition: A deep survey.
\newblock {\em Expert Systems with Applications}, 164:113794, 2021.

\bibitem{rotoli2022emergency}
Jason~M Rotoli, Sarah Hancock, Chanjun Park, Susan Demers-Mcletchie, Tiffany~L Panko, Trevor Halle, Jennifer Wills, Julie Scarpino, Johannah Merrill, Jeremy Cushman, et~al.
\newblock Emergency medical services communication barriers and the deaf american sign language user.
\newblock {\em Prehospital emergency care}, 26(3):437--445, 2022.

\bibitem{rust2024towards}
Phillip Rust, Bowen Shi, Skyler Wang, Necati~Cihan Camg{\"o}z, and Jean Maillard.
\newblock Towards privacy-aware sign language translation at scale.
\newblock {\em arXiv preprint arXiv:2402.09611}, 2024.

\bibitem{schnepp2020human}
Jerry Schnepp, Rosalee Wolfe, Gilbert Brionez, Souad Baowidan, Ronan Johnson, and John McDonald.
\newblock Human-centered design for a sign language learning application.
\newblock In {\em Proceedings of the 13th ACM International Conference on PErvasive Technologies Related to Assistive Environments}, PETRA '20, New York, NY, USA, 2020. Association for Computing Machinery.

\bibitem{schoenherr2024ai}
Jordan~Richard Schoenherr and Robert Thomson.
\newblock When ai fails, who do we blame? attributing responsibility in human-ai interactions.
\newblock {\em IEEE Transactions on Technology and Society}, 2024.

\bibitem{slinto}
ShuR.
\newblock Slintodictionary, 2021.

\bibitem{NEURIPS2023_00dada60}
Thad Starner, Sean Forbes, Matthew So, David Martin, Rohit Sridhar, Gururaj Deshpande, Sam Sepah, Sahir Shahryar, Khushi Bhardwaj, Tyler Kwok, Daksh Sehgal, Saad Hassan, Bill Neubauer, Sofia Vempala, Alec Tan, Jocelyn Heath, Unnathi Kumar, Priyanka Mosur, Tavenner Hall, Rajandeep Singh, Christopher Cui, Glenn Cameron, Sohier Dane, and Garrett Tanzer.
\newblock Popsign asl v1.0: An isolated american sign language dataset collected via smartphones.
\newblock In A.~Oh, T.~Neumann, A.~Globerson, K.~Saenko, M.~Hardt, and S.~Levine, editors, {\em Advances in Neural Information Processing Systems}, volume~36, pages 184--196. Curran Associates, Inc., 2023.

\bibitem{terry2024scoping}
Julia Terry and Rhian Meara.
\newblock A scoping review of deaf awareness programs in health professional education.
\newblock {\em PLOS Global Public Health}, 4(8):e0002818, 2024.

\bibitem{9413126}
Elahe Vahdani, Longlong Jing, Yingli Tian, and Matt Huenerfauth.
\newblock Recognizing american sign language nonmanual signal grammar errors in continuous videos.
\newblock In {\em 2020 25th International Conference on Pattern Recognition (ICPR)}, pages 1--8, 2021.

\bibitem{10.1145/3544548.3581278}
Qiaosi Wang, Michael Madaio, Shaun Kane, Shivani Kapania, Michael Terry, and Lauren Wilcox.
\newblock Designing responsible ai: Adaptations of ux practice to meet responsible ai challenges.
\newblock In {\em Proceedings of the 2023 CHI Conference on Human Factors in Computing Systems}, CHI '23, New York, NY, USA, 2023. Association for Computing Machinery.

\bibitem{weaver2011we}
Kimberly~A. Weaver and Thad Starner.
\newblock We need to communicate! helping hearing parents of deaf children learn american sign language.
\newblock In {\em The Proceedings of the 13th International ACM SIGACCESS Conference on Computers and Accessibility}, ASSETS '11, page 91–98, New York, NY, USA, 2011. Association for Computing Machinery.

\bibitem{wilcox1994multimedia}
S.~Wilcox, J.~Scheibman, D.~Wood, D.~Cokely, and W.~C. Stokoe.
\newblock Multimedia dictionary of american sign language.
\newblock In {\em Proceedings of the First Annual ACM Conference on Assistive Technologies}, Assets '94, page 9–16, New York, NY, USA, 1994. Association for Computing Machinery.

\bibitem{10.1145/2907069}
Jacob~O. Wobbrock and Julie~A. Kientz.
\newblock Research contributions in human-computer interaction.
\newblock {\em Interactions}, 23(3):38–44, apr 2016.

\bibitem{wong2010interactive}
Rita Wong, Norman Poh, Josef Kittler, and David Fr{\"o}hlich.
\newblock Interactive quality-driven feedback for biometric systems.
\newblock In {\em 2010 Fourth IEEE International Conference on Biometrics: Theory, Applications and Systems (BTAS)}, pages 1--7. IEEE, 2010.

\bibitem{yanovich2016detection}
Polina Yanovich, Carol Neidle, and Dimitris Metaxas.
\newblock Detection of major {ASL} sign types in continuous signing for {ASL} recognition.
\newblock In {\em Proceedings of the Tenth International Conference on Language Resources and Evaluation ({LREC}'16)}, pages 3067--3073, Portoro{\v{z}}, Slovenia, May 2016. European Language Resources Association (ELRA).

\end{thebibliography}

\appendix
\newpage

\section{Narratives}
\label{app:narratives}

\textbf{Narrative 1:} I loved to visit with my uncle while I was growing up.  We often went to the zoo and saw a lot of different animals, tigers, lions, penguins etc.  My favorite part about my visits was because my uncle was a skilled cook.  He made the best chicken spaghetti I have ever tasted.  I think he learned to cook from his grandparents who immigrated from Italy, or maybe from the other side of the family who immigrated from France.  I remember no matter the family event, wedding, funeral, new baby birth, the whole family would get together and eat delicious food along with salad, garlic bread and lots of desserts like cookies, cakes and pies.  We ate until we were satisfied and made a great memory. 

\textbf{Narrative 2:} My grandparents’ farm was a favorite place for me to visit growing up. Grandma and Grandpa grew up in Germany where their family had a farm.  After moving to the U.S. they decided to begin raising and selling all kinds of animals.  We had to tend to chickens, goats, ducks, and pigs.  Grandpa also had two donkeys that were the cutest! They ate oats and hay. Every day at lunchtime Grandma made us sandwiches and soup that were delicious.  

\textbf{Narrative 3:} I was so excited to plan my upcoming trip! I made plans for a 2-month long trip.  I made arrangements for someone to care for my dogs, horses and cows. On the morning of my departure I ate a good breakfast of eggs, bacon, toast and coffee because I knew it was a long journey.  I arrived at the airport, boarded my plane and then flew for 25 hours!  Upon disembarking, the weather was hot and humid, but I had arrived in beautiful Cambodia!  What an incredible place!  The people were so friendly, and kind and I enjoyed sightseeing. I then took a boat ride to Vietnam.  There I learned a lot about the history shared between the U.S. and Vietnam.  Then I traveled to China, but upon my arrival realized I had lost my passport!  I had left it back in Cambodia and was panicked! Fortunately, the hotel back in Cambodia was able to send it to me.  On my Chinese adventure I tried lots of new things, delicious food and even ate bugs! What an amazing trip.  

\textbf{Narrative 4:} What a beautiful sunny day!  I think I will call some friends and go to the zoo.  There are so many animals to see.  When I was a child my favorite animals to see were apes, specifically, the orangutan.  They were always so playful and often had new babies that they cared for.  Today, I hope to visit the giraffes, tigers, lions and the new baby elephant.  The zoo plans to have a contest to name the baby.  People will send in ideas from across the state and the best one will be chosen.  I suggest they name him Jazz since he already has a trumpet!

\newpage

\onecolumn

\section{User Interface}
\label{app:user_interface_figs}

\begin{figure}[H]
    \centering
    \includegraphics[width=0.8\linewidth]{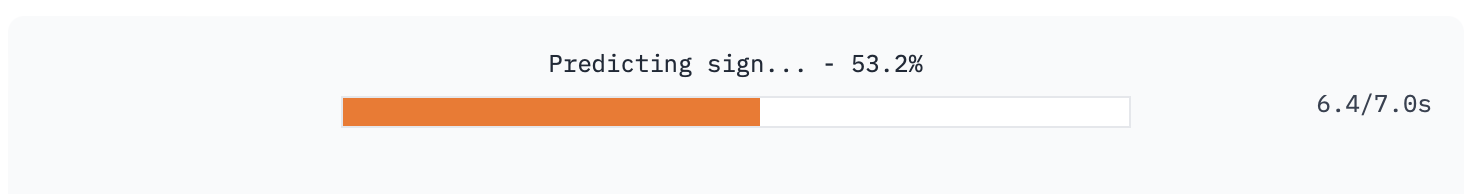}
    \caption{Progress bar showing the live progress of the recognition and an estimated time remaining.}
    \Description[Recognition progress bar]{An orange-shaded region covers one-half of the white-background progress bar. Above this progress bar is a ‘Predicting sign… - 53.2\%’ label. To the right of this progress bar is a label saying ‘6.4/7.0s’.}
    \label{fig:progressbar}
\end{figure}

\begin{figure}[H]
    \centering
    \subfigure[Example warning message]{\centering\includegraphics[width=0.4\linewidth]{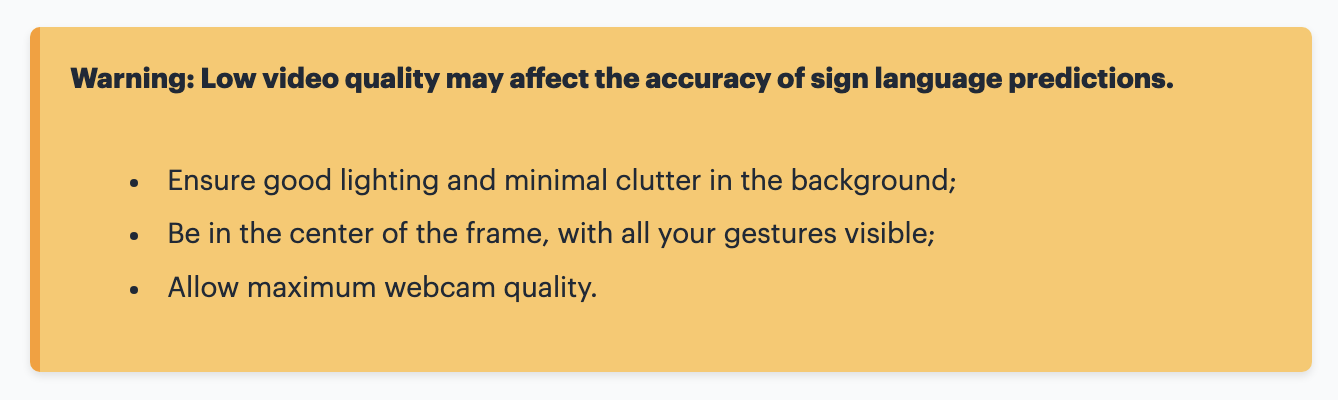}}
    \Description[Warning message box]{The yellow-background box has a bold warning label at the top and three regular-font bullet points below.}
    \subfigure[Example error message]{\centering\includegraphics[width=0.4\linewidth]{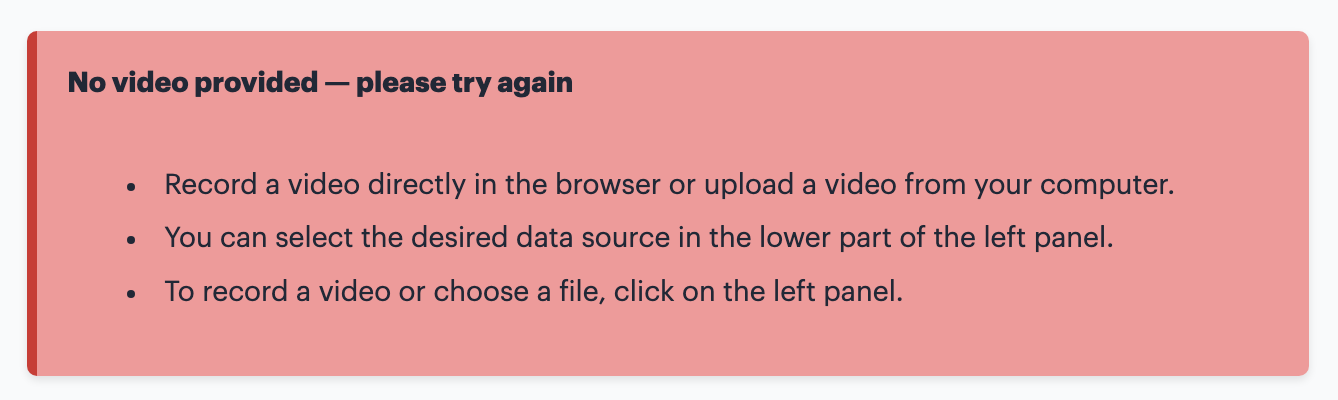}}
    \Description[Error message box]{The red-background box has a bold warning label at the top and three regular-font bullet points below.}
    \caption{Example message boxes presented after the user submits an input video for analysis if problematic features are detected in the input video. The bold text contains a summary of the problem; the bullet points present suggestions for fixing the problem.}
    \label{fig:errorwarningmessage}
\end{figure}

\FloatBarrier

\begin{figure}[H]
    \centering
    \includegraphics[width=0.4\linewidth]{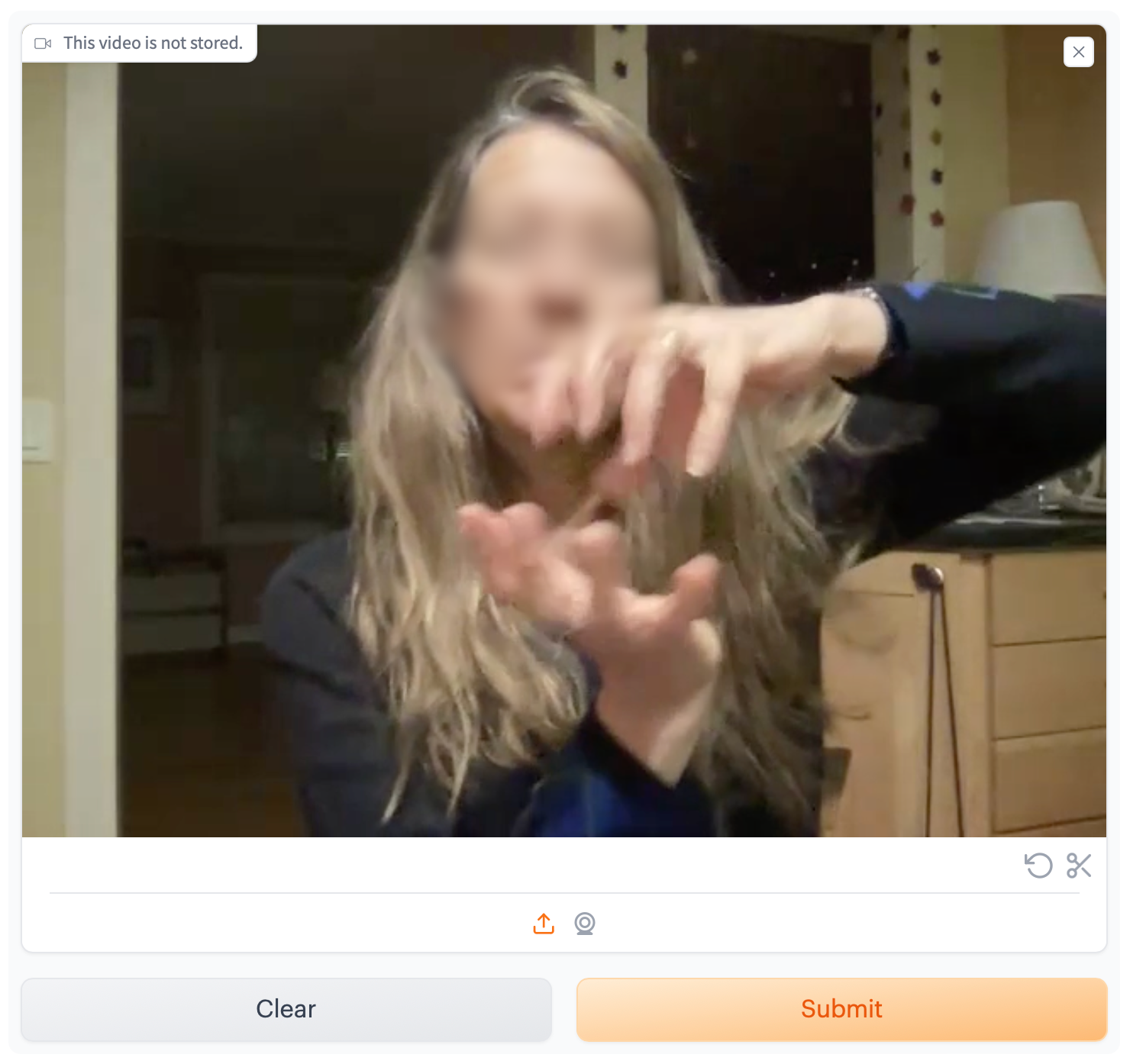}
    \Description[Video preview after upload]{The video uploading and recording interface is composed of a preview box and two buttons. A blurred-out frame of a person signing is shown in the box at the top. On the bottom right of this box is a loop and a scissors icon, representing the buttons to reset the trimming or initiate the trimming, respectively. On the bottom center of this box is an arrow pointing up and a webcam icon representing the data source of the uploaded video. Below this box are two buttons: on the left, a gray ‘Clear’ button, and on the right, an orange ‘Submit’ button.}
    \caption{The dictionary supports live webcam recording and file uploads (icons indicate the data source). The `Clear' button removes the video and the `Submit' button starts the prediction.}
    \label{fig:videopreview-basic}
\end{figure}

\begin{figure}[H]
    \centering
    \includegraphics[width=0.35\linewidth]{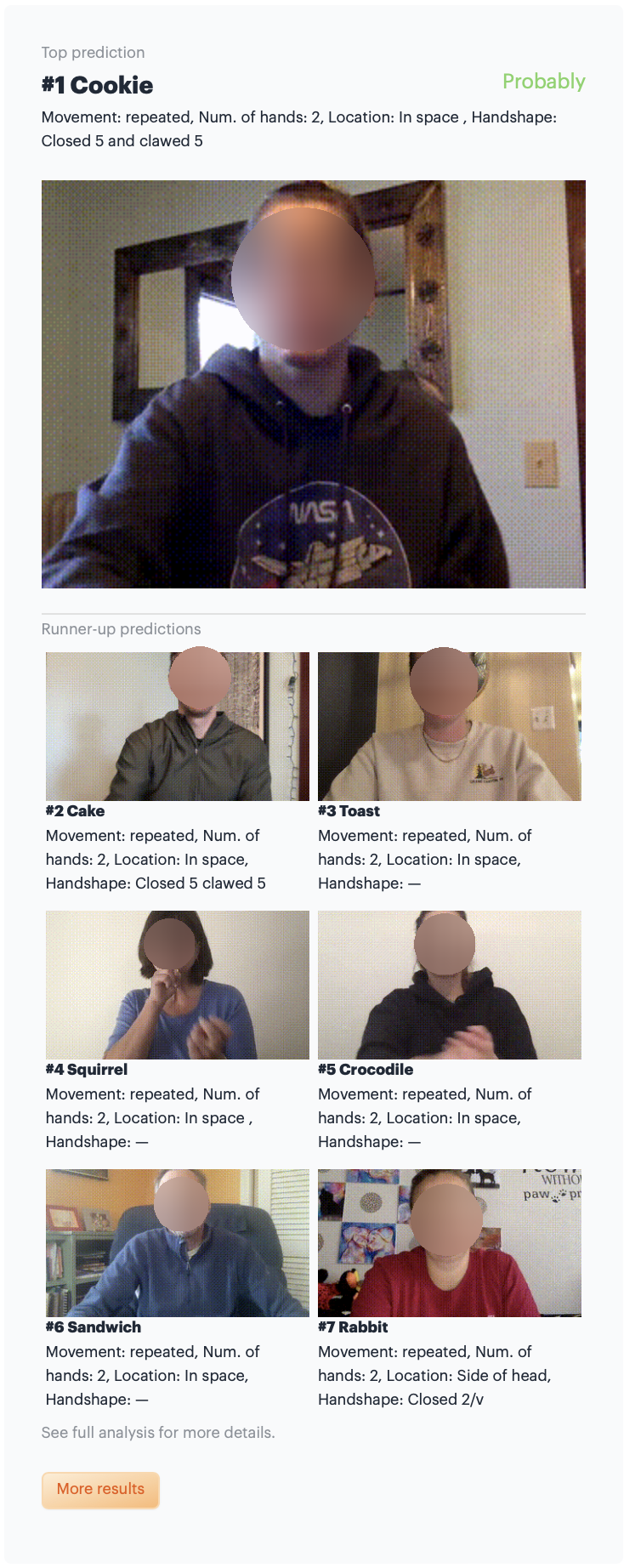}
    \caption{The compact result view on the `Main dictionary interface' page presents the model's top $7$ predictions. The top-$1$ prediction is highlighted in a larger size at the top, while the remaining $6$ predictions appear below in smaller size. Each entry includes a representative recording, word translation, probability score, and metadata.}
    \Description['Main dictionary interface' results view]{The vertical view with a gray background shows the top six prediction results. The single top prediction, presented at the top of the view, shows the word translation of the sign (‘Cookie’) on the left and its textual probability on the right (‘Probably’). Below these labels is a set of metadata associated with the sign. A blurred-out frame of a video with a person signing the sign is below all these labels, covering the entire width of the view. The remaining six predictions are presented in a 3x2 grid, each covering one-half of the view width. Each record presents the word translation of the sign, the metadata, and the blurred-out frame of a video with a person signing the respective sign. An orange ‘More results’ view is located at the very bottom of this view.}
    \label{fig:resultsview-main}
\end{figure}

\end{document}